\documentclass[aip,graphicx]{revtex4-1}
\usepackage{amsfonts}
\usepackage{graphicx}% Include figure files
\usepackage{dcolumn}% Align table columns on decimal point
\usepackage{bm}% bold math

\usepackage[utf8]{inputenc}
\usepackage[T1]{fontenc}
\usepackage{mathptmx}
\usepackage{etoolbox}
\usepackage{float}
\usepackage{amsmath}
\usepackage[utf8]{inputenc}
\draft % marks overfull lines with a black rule on the right

\begin{document}

%\preprint{}

\title{Investigation of dust grains by optical tweezers for space applications} %Title of paper



\author{A. Magazz\`u}
\affiliation{CNR-IPCF, Istituto per i Processi Chimico-Fisici, Messina, Italy}

\author{D. Bronte Ciriza}
\affiliation{CNR-IPCF, Istituto per i Processi Chimico-Fisici, Messina, Italy}
\affiliation{Dipartimento di Scienze Matematiche e Informatiche, Scienze Fisiche e Scienze della Terra, Universit\`a di Messina, Italy}

\author{A. Musolino}
\affiliation{Dipartimento di Scienze della Terra, Universit\`a di Pisa, Pisa, Italy}
\affiliation{ INAF-IAPS, Istituto Nazionale di Astrofisica, Istituto di Astrofisica e Planetologia Spaziali, Rome, Italy}

\author{A. Saidi}
\affiliation{Dipartimento di Scienze Matematiche e Informatiche, Scienze Fisiche e Scienze della Terra, Universit\`a di Messina, Italy}

\author{P. Polimeno}
\affiliation{CNR-IPCF, Istituto per i Processi Chimico-Fisici, Messina, Italy}
\affiliation{Dipartimento di Scienze Matematiche e Informatiche, Scienze Fisiche e Scienze della Terra, Universit\`a di Messina, Italy}

\author{M. G. Donato}
\affiliation{CNR-IPCF, Istituto per i Processi Chimico-Fisici, Messina, Italy}

\author{A. Foti}
\affiliation{CNR-IPCF, Istituto per i Processi Chimico-Fisici, Messina, Italy}

\author{P. G. Gucciardi}
\affiliation{CNR-IPCF, Istituto per i Processi Chimico-Fisici, Messina, Italy}

\author {M. A. Iat\`\i}
\affiliation{CNR-IPCF, Istituto per i Processi Chimico-Fisici, Messina, Italy}

\author{R. Saija}
\affiliation{Dipartimento di Scienze Matematiche e Informatiche, Scienze Fisiche e Scienze della Terra, Universit\`a di Messina, Italy}

\author{N. Perchiazzi}
\affiliation{Dipartimento di Scienze della Terra, Universit\`a di Pisa, Pisa, Italy}

\author{A. Rotundi}
\affiliation{Dipartimento di Scienze e Tecnologie, Universit\`a degli studi di Napoli Parthenope, Italy}

\author{L. Folco}
\affiliation{Dipartimento di Scienze della Terra, Universit\`a di Pisa, Pisa, Italy}
\affiliation{CISUP, Center for Instrument Sharing of the University of Pisa, Universit\`a di Pisa, Italy}

\author{O. M. Marag\`{o}}
\affiliation{CNR-IPCF, Istituto per i Processi Chimico-Fisici, Messina, Italy}


\date{\today}

\begin{abstract}
Cosmic dust plays a dominant role in the universe, especially in the formation of stars and planetary systems. Furthermore, the surface of cosmic dust grains is the bench-work where molecular hydrogen and simple organic compounds are formed.
We manipulate individual dust particles in water solution by contactless and non-invasive techniques such as standard and Raman tweezers, to characterize their response to mechanical effects of light (optical forces and torques) and to determine their mineral compositions. Moreover, we show accurate optical force calculations in the T-matrix formalism highlighting the key role of composition and complex morphology in optical trapping of cosmic dust particles.
This opens perspectives for future applications of optical tweezers in curation facilities for sample return missions or in extraterrestrial environments.

\end{abstract}


%% We recommend that authors also use the natbib \citepp
%% and \citept commands to identify citations.  The citations are

\maketitle %\maketitle must follow title,

\section{Introduction} \label{sec:intro}

In the frame of astrophysical science and its related technology (space missions, optical and radio telescopes), cosmic dust has attracted the interest of scientists due to its role in the cycling processes active in the universe. Cosmic dust is a comprehensive term indicating small solid particles with sizes ranging from a few nanometers to tenths of millimeter. They are floating around in the interstellar medium or in the interplanetary space in the solar system. %from the nanometer to millimeter range moving in the interstellar medium in molecular clouds or in the interplanetary space in the solar system.
Interstellar dust is mainly generated by the lifecycles of many generations of stars: it is released by radiation pressure and solar wind or ejected during the end-time explosion of stars or during the blowing off of their outer layers \citep{calura2008cycle, woosley2002evolution, woosley1995evolution}. Interplanetary dust consists of small solid particles generated by collisions between solid bodies (e.g asteroids, planets and their satellites) or evaporation of icy bodies (e.g. comets \citep{rietmeijer1998interplanetary}). If interstellar dust is almost exclusively analysed through remote observations \citep{draine2003interstellar,loddersamari2005presolar}, on the other hand interplanetary dust is available through sample-return space missions from interplanetary medium, planets and minor bodies \citep{westphal2014evidence, brownlee2003stardust, frank2014stardust},  from the Earths' stratosphere \citep{testa1990collection, taylor1996discovery, della2012insitu, lauretta2017osiris} and at the Earth's surface in the form of micrometeorites  (e.g., \citep{genge2008classification, folco2015micrometeorites, taylor2016cosmic}). Samples are analysed at terrestrial facilities by the state-of-the-art analytical techniques – since the size of some instruments is still too large to fly to the space  \citep{rietmeijer2001identification, rauf2010evidence, della2014introducing}. 

Physico-chemical properties of cosmic dust can be studied by different techniques, such as X-rays diffraction \citep{mackinnon1987mineralogy}, SEM, TEM \citep{lewis1987interstellar}, IR, Raman \citep{rotundi2008combined, davidson2012nanosims}, mass spectroscopy \citep{floss2006identification}. However, these techniques may induce shielding effects by the substrates or by other particles. In this work, to overcome these unwanted effects, we develop and test our setups of optical and Raman tweezers, that are contactless and non-destructive techniques useful for the manipulation and investigation of individual grains of cosmic dust \citep{alali2020laser}. % abbas2007lunar
Optical tweezers (OT) \citep{jones2015optical} are tools based on focused laser beams, that allow the trapping and manipulation of micro and nanoparticles without a physical contact. This contactless and non-destructive technique in its simplest configuration is based on a single laser beam focused down to the diffraction limit by a high numerical aperture objective \citep{ashkin1971optical, ashkin1986observation}. 
Optical forces are strongly dependent on particle size, shape and composition (see also Section \ref{AOF}). However, OT enable trapping and characterization of particles in a wide size range, from single atoms to cells \citep{jones2015optical, polimeno2018optical}. Furthermore, Raman tweezers, i.e. OT coupled to a Raman spectrometer, are able to spectroscopically characterize optically trapped samples allowing accurate information on their composition \citep{thurn1984raman, lankers1994raman, pan2012photophoretic, gong2018optical, gillibert2019raman}.

Here, we characterize the dust particle response to optical forces and torques, providing accurate information on the particle dynamics in the optical trap. Furthermore, we identify dust particle mineralogical compositions by Raman tweezers. We compare our findings with optical force calculations based on the T-matrix formalism showing how the composition and the complex shape of dust particles play a key role in their light-driven dynamics.

\section{Materials and Methods}

The samples trapped and characterized with our Raman tweezers have known textures and mineralogic compositions documented in this work and in previous studies. The aim is to reproduce the well-known composition of the  trapped particles to validate the application of optical manipulation techniques to cosmic dust.

\begin{figure}
  \centering
  \resizebox{12 cm}{!}{%
  \includegraphics{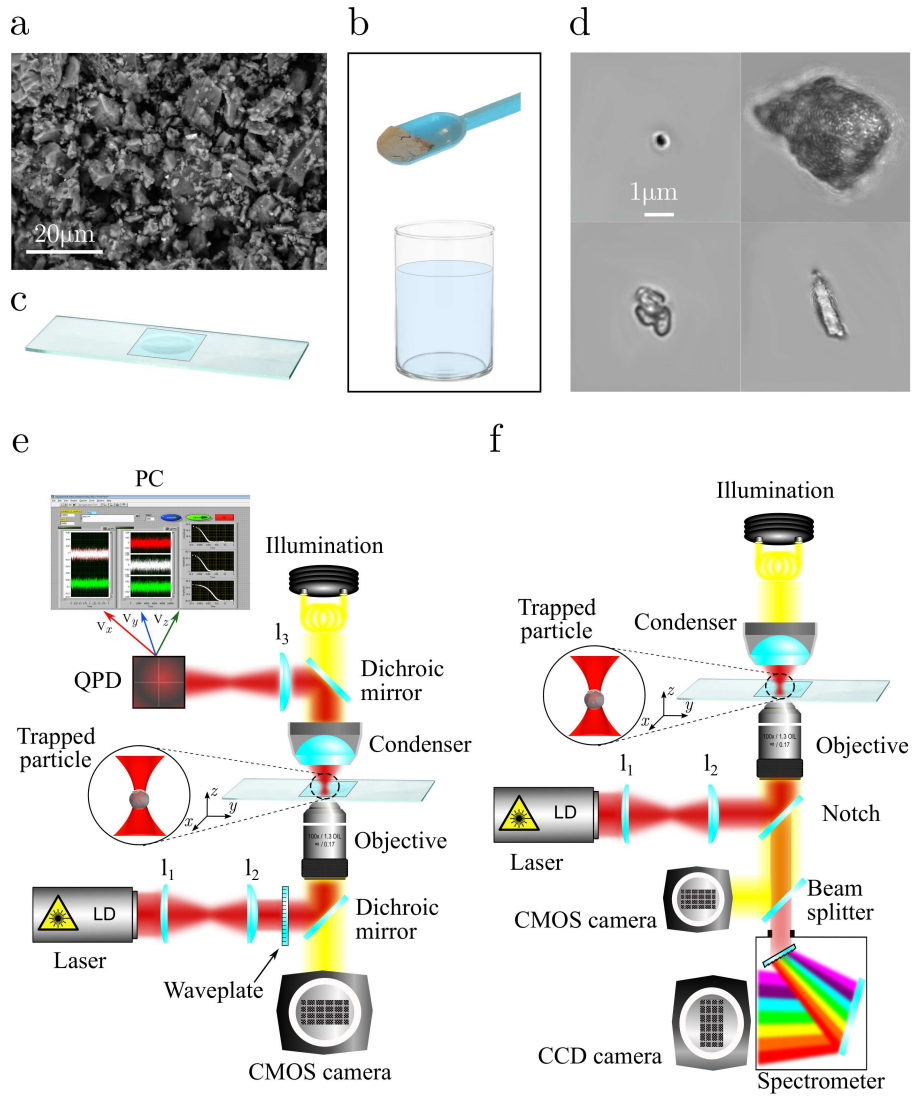}
  }
  \caption{Samples preparation and sketch of the experimental setups. a) SEM image of DEW 12007 lunar meteorite powdered sample, showing the poly-crystalline nature of each individual grains. b) Dust samples are dispersed in distilled water by ultrasound sonication and c) this solution is placed within a cavity glass slide and sealed by a coverglass. d) Screenshots of four different 3D optically trapped grains of cosmic dust in water solution having different size and morphology.
   e)  Standard optical tweezers setup. f) Raman tweezers setup.}
  \label{fig1}
\end{figure}

The samples used have terrestrial and extraterrestrial origins. They are (1) a quartzarenite from Kamil Crater in Egypt (M26) \citep{fazio2014shock}, (2) a hawaiite from Etna volcano in Italy (HE-1), (3) the CV3-OxA carbonaceous chondrite Allende (A-1), and (4) a lunar meteorite (polymict regolith breccia) found in Antarctica (DEW 12007) \citep{collareta2016high}. Previously, mineral grains used to calibrate the dust analyser Giada on board of Rosetta space mission \citep{colangeli2007grain} were used as the simplest possible starting material (monomineralic, uniform size range). In this work, we have chosen more complex non-uniform samples: the terrestrial rocks M-26 and HE-1, considered inasmuch analogues of planetary materials (mono and polymineralic); and the extraterrestrial rocks as representative of primitive (A-1) and differentiated (DEW 12007) bodies of the solar system.  

Once powdered, the samples are analysed using microanalytical scanning electron microscopy for a textural characterization (shape, grain size). Their main mineralogy is determined using X-ray powder diffraction (Section~\ref{ASC}). The characterization was conducted at the Dipartimento di Scienze della Terra and at the Center for Instrument Sharing of the University of Pisa (CISUP). 
 
All the analysed samples were provided as dried powder, whose grains showed a non-homogeneous size and shape distribution, as showed in Fig. \ref{fig1}(a,d).  The first step of the sample preparation was the dispersion of the powder in distilled water by ultrasound sonication with an appropriate concentration suitable for optical trapping (e.g. few particles / microliter). The water-dust solution was then placed in a glass cavity slide and sealed by a coverglass, as showed in Fig. \ref{fig1}(b,c).  The sealed cavity glass slide was then placed on the sample holder of our optical tweezers for the investigation Fig.  1(c).
For the investigation of individual grain dust particles dispersed in water we used two different customised setups: i) standard OT, where an optical trap was generated by highly focusing a laser beam through an objective, the trajectory of the trapped particle was acquired by a quadrant photodiode (QPD) and analysed by a computer to calculate optical forces and rotations arising from the interaction of the particle with light, Fig. \ref{fig1}(e); ii) Raman tweezers, where the back scattered light from the particle was collected through the same focusing objective and reflected to a Raman spectrometer for phase identification, Fig. \ref{fig1}(f). See Section \ref{AES} for more details.

\section{Results}
\subsection{Optical forces and torques on dust particles}
\begin{figure}
 \centering
  \resizebox{14 cm}{!}{%
  \includegraphics{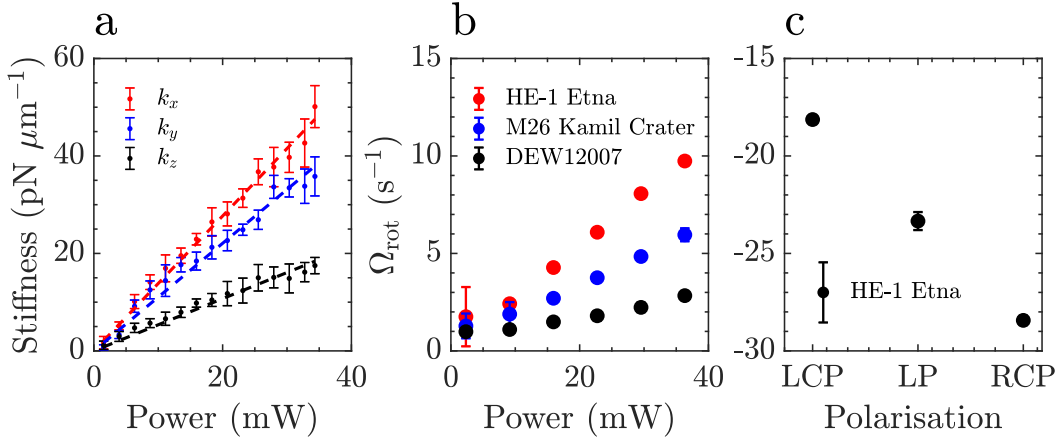}
  }
  \caption{Effects of the laser light  on  trapped particles. a) Trap stiffnesses $\kappa_x$ , $\kappa_y$ and $\kappa_z$ as a function of the laser power measured at the objective for a single trapped grain of the lunar meteorite DEW 12007. b) Rotational frequencies $\Omega_{\rm rot}$ of different samples as a function of the laser power measured at the objective. c) Rotational frequencies, $\Omega_{\rm rot}$, of an optically trapped single dust grains of HE-1  for  different polarised light: left circular polarised (LCP), linear polarised (LP) and right circular polarised (RCP) light. Negative/positive frequency values correspond to a rotation anticlockwise/clockwise with respect to the z axis, respectively.}
  \label{fig2}
\end{figure}

The output signals from the QPD, proportional to the particle displacement of the trapped particle from its equilibrium position, are analysed by MATLAB routines to calculate the trap stiffness and to detect any rotation of the particle in the perpendicular plane $x-y$. In particular, we use the autocorrelation functions (ACFs) and power spectral density (PSD) calibration methods \citep{jones2015optical, gieseler2021optical} to obtain the relaxation frequencies of the trapped particles as fitted parameters and hence to calculate the trap stiffness, Figs. \ref{apx}(b), \ref{fig2}(a). See Section \ref{AES} for more details. We observe that the stiffness of terrestrial and extra-terrestrial samples, optically trapped by linear polarised light, increases with the laser power, as showed in Fig. \ref{fig2}(a) for a single trapped grain of the lunar meteorite DEW 12007. Here it is possible to notice that the stiffnesses $\kappa_x$, $\kappa_y$ and $\kappa_z$ increase almost linearly with the laser power and the difference between the stiffnesses along the $x$ and $y$ directions can be due to a possible asymmetry of the trapped grain with respect to the propagation axis $z$. Usually the values of the stiffness $\kappa_z$ are lower than the values of $\kappa_x$ and $\kappa_y$ due to the longer extension of the gaussian beam along the $z$ axis \citep{jones2015optical}.
Due to the linearity of the stiffness with power it is possible to obtain the stiffness efficiency $q_{i}= k_{i}/\rm {Pw}$ (with $i=x,y,z$) as fitting parameter from the values of $\kappa_i$ reported in Fig. \ref{fig2}(a), where $\rm {Pw}$ is the value of laser power measured at the objective output by a power meter. In particular, by a linear fit, we obtain that $q_x=1.38$ ${\rm pN}\mu {\rm m}^{-1}{\rm mW}^{-1}$, $q_y=1.104$ ${\rm pN}\mu {\rm m}^{-1}{\rm mW}^{-1}$ and $q_z=0.536$ ${\rm pN}\mu {\rm m}^{-1}{\rm mW}^{-1}$ respectively. %A possible asymmetry of the particle can also make the trapping process more or less efficient along a specific direction, such as in the case of the lunar meteorite DEW 12007 where $q_x>q_y$. 
Radiation pressure on complex shaped or absorbing particles can also have a destabilising effect by pushing the trapped particles in regions of lower intensity along the $z$ axis. For example, grains from Allende meteorite are difficult to trap in 3D and were often investigated while trapped in 2D against the cell wall.

\begin{figure}
  \centering
  \resizebox{14 cm}{!}{%
  \includegraphics{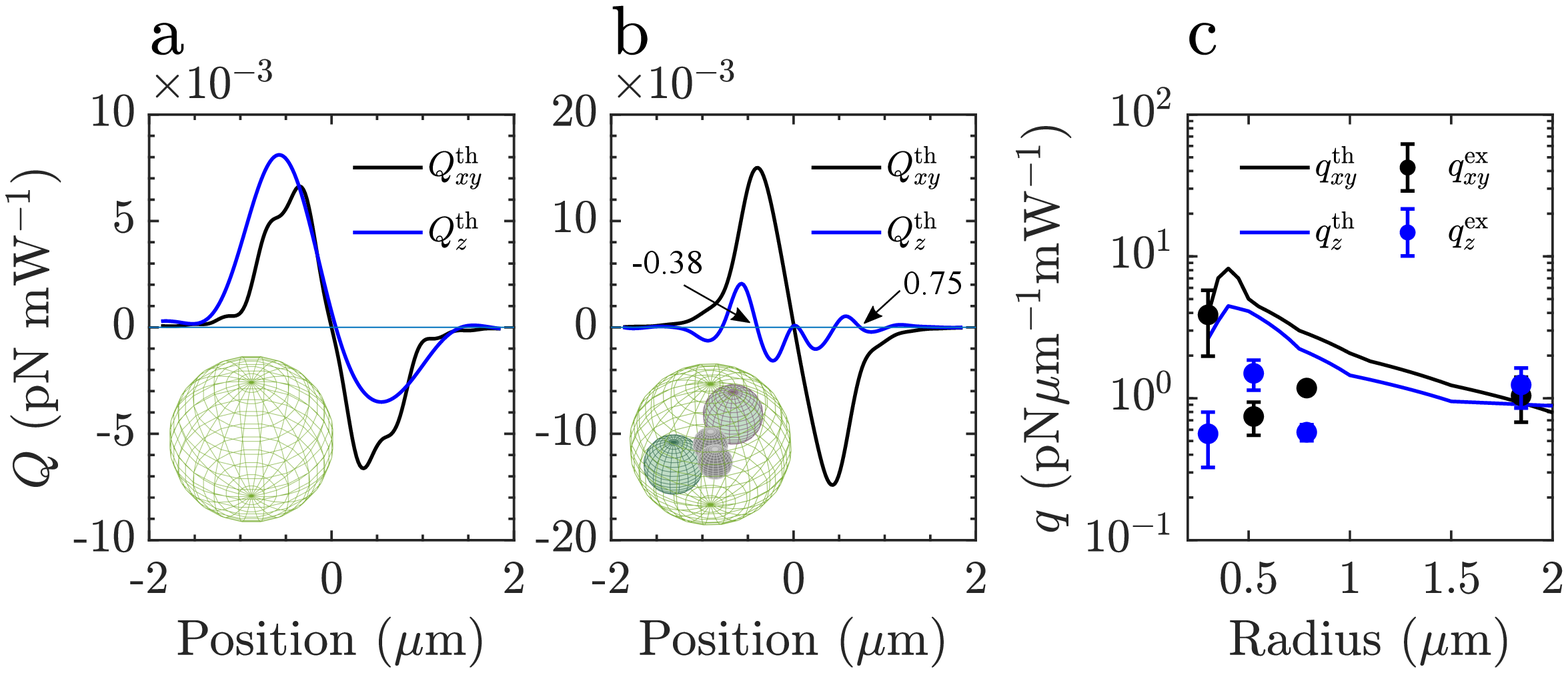}
  }
  \caption{Theoretical trap efficiencies.  a) Black line represents the theoretical trap efficiency $Q_z^{\rm th}$ along the longitudinal direction $z$ with $x=y=0$. Dark blue line represents the theoretical trap efficiency $Q_{xy}^{\rm th}$ along the transversal direction $x-y$ with $z=0$. In the inset is showed the theoretical particle, modelled as a micro-sphere with homogeneous refractive index according to the Bruggeman mixing rule. Light blue line is a reference line at $y=0$. b) Black line represents the theoretical trap efficiency $Q_z^{\rm th}$ for a particle model showed in the inset. Here the particle is modelled again as a micro-sphere with homogeneous refractive index according to the Bruggeman mixing rule but has 4 inclusions made of the secondary constituents of the lunar meteorite,  $2$ inclusions made of olivine having a $10\%$ each of the total volume and $2$ inclusions of ilmenite with a $2\%$ each of the volume. $Q_z^{\rm th}$ presents only two stable equilibrium points at $x=-0.38$ and $x=0.75\ \mu {\rm m}$, but, according to our forces calculations, only at $x=0.75\ \mu {\rm m}$ trapping along the $x$ and $y$ direction is possible.
  c) Theoretical (lines) and experimental (dots) stiffness efficiencies $q_{xy}$ and $q_z$ within the transversal plane $x-y$ and along the longitudinal direction $z$ respectively. Blue lines and dots represent the theoretical and experimental efficiency  $q_{xy}^{\rm th}$ and $q_{xy}^{\rm ex}$ respectively, along the transversal direction $x-y$, while the black lines and dots represent the theoretical and experimental efficiency  $q_{z}^{\rm th}$ and $q_{z}^{\rm ex}$ respectively, along the longitudinal direction $z$.}
  \label{fig3}
\end{figure}

In order to compare our experimental results with theoretical expectations, we calculate optical forces through the Maxwell stress tensor exploiting the multipole expansion and T-matrix formalism \citep{borghese2007scattering, polimeno2018optical, polimeno2021optical}. We consider particle models emulating the dust grains of the lunar meteorite DEW 12007. Although all the trapped dust grains showed irregular shapes, for simplicity, we modelled our particles as micro-spheres having an average diameter $d=1\ \mu \rm{m}$ with a refractive index given by the Bruggeman mixing rule for the constituent minerals of DEW 12007, as showed in the inset of figure \ref{fig3}(a)   \citep{bohren2008absorption,polimeno2021optical}. 

In our calculations we consider a single Gaussian laser beam  having a wavelength $\lambda = 830\ \rm{nm}$ and a power $\rm {Pw}=50\ \rm{mW}$, focused by a high-NA objective (NA=1.3), mimicking the experimental conditions used during the investigation of the lunar dust grains. In figure \ref{fig3}(a), we show the theoretical transversal and longitudinal trap efficiencies defined respectively as: $Q_{xy}^{\rm th}= (F_x^{\rm opt}+F_y^{\rm opt})/2\rm {Pw}$ and $Q_z^{\rm th}=F_z^{\rm opt}/\rm {Pw}$. Thereafter, we refined our particle model to take into account the anisotropy and heterogeneity of the real dust grains. We still considered an homogeneous sphere, with a refractive index obtained according to the Bruggeman mixing rule but now we add $4$ spherical inclusions, \ref{fig3}(b) inset. These inclusions are made of the secondary constituents of the lunar meteorite, in particular we consider $2$ inclusions made of olivine having a $10\%$ each of the total volume of the modelled grain, and $2$ inclusions of ilmenite with a $2\%$ each of the total volume. Similarly to the previous model, we calculate the optical forces and the theoretical longitudinal trap efficiency $Q_z^{\rm th}$, as showed in Fig. \ref{fig3}(b). Here it is possible to notice how the trap efficiency is affected by the internal structure of a dust grain, presenting several equilibrium points with $Q_z^{\rm th}=0$, and only two stable equilibrium points at $x=-0.38$ and $x=0.75\ \mu {\rm m}$, conversely, the homogeneous model was showing an only stable equilibrium point at $x=0\ \mu {\rm m}$, Fig. \ref{fig3}(a,b).   

%After computing the theoretical optical forces acting on our model particles and their stiffnesses, % we compare the theoretical and experimental stiffnesses to validate our theoretical model.   %we get the theoretical trap stiffnesses normalized to incident power and compare them to the values measured in our experiments.

Aiming at validating our theoretical model, we compare the theoretical $q^{\rm th}_{xy,z}$ and experimental $q^{\rm ex}_{xy,z}$ stiffness efficiencies on the transverse $x-y$ plane and in the longitudinal direction $z$. In particular, in the former case the mean transverse efficiency $q_{xy}= (q_x+q_y)/2$ is considered. Moreover, in the case of the theoretical stiffness efficiencies, $\rm {Pw}$ is the value of power set for the simulation.

%\textbf{we compare the theoretical and experimental stiffness efficiencies $q^{\rm th /ex}$ and $q^{\rm ex}$ in the longitudinal direction $q_z$ and in the transverse plane $q_{xy}= (q_x+q_y)/2$, Fig.\ref{fig3}(c), to validate our theoretical model. In the case of the theoretical stiffness efficiencies, $\rm {Pw}$ is the value of power set for the simulation.}

In figure \ref{fig3}c it is possible to notice a relative good agreement between the theoretical and experimental values of the stiffness efficiencies  $q^{\rm th}$ and $q^{\rm ex}$, validating the theoretical model used for the calculation of optical forces acting on cosmic dust. The small discrepancies are due to the spherical shape used in our model which is a simplification of the complex particle shape, mass distribution, and mineral composition. %Due to the difficulties  of refining the model by knowing the exact geometry, distribution and mineral composition of the optically trapped grains.
%It is noteworthy that the shape of the particles can play an important role in the light - particle interaction. 
In fact, a linearly polarised laser light does not produce any radiation torque on a spherical and homogeneous particle because of symmetry  \citep{marston1984radiation}. However, for anisotropic and asymmetric particles two additional mechanical effects of light occur: i) a transverse component of radiation pressure which is responsible for the \textit{optical lift effect}, i.e., a transverse displacement of the particle with respect to the incident light propagation direction \citep{swartzlander2011stable}; ii) a radiation pressure torque inducing particle rotations and known as \textit{windmill effect} \citep{jones2015optical} %the radiation pressure has a transverse component, which is responsible for the optical lift effect, e.g.: a transversal displacement of the particle with respect to the incident light propagation direction \citep{jones2015optical}.% For more complex shapes, multiple events of light scattering can occur within the particle, producing a radiation torque on the particle and inducing their rotation. This phenomenon is known as windmill effect \citep{swartzlander2011stable}.

Rotations of a trapped particle induce a correlation among its $x$, $y$ trajectories. In these circumstances, we can use optical tweezers also for \textit{photonic torque microscopy}  to quantify these rotations \citep{irrera2016photonic, schmidt2018microscopic}. In particular, particle rotations in the $x-y$ plane  can be highlighted by calculating the differential cross correlations ${\rm DCCF}_{xy}(\tau)=\left \langle x(t)y(t+\tau) \right \rangle-\left \langle y(t)x(t+\tau) \right \rangle$ of the acquired signal from the QPD, Fig. \ref{apx}(a), for different lag times  $\tau$. The rotational frequency $\Omega_{\rm rot}$ can be obtained as fitting parameter of the ${\rm DCCF}_{xy}$ by a sinusoidal model \citep{pesce2009quantitative,jones2009rotation}.

In Fig. \ref{fig2}(b) we show the rotational frequencies $\Omega_{\rm rot}$ of three different trapped samples, where $\Omega_{\rm rot}$ increases as the laser power increases. It is noteworthy that the polarisation used to trap the samples listed in Fig. \ref{fig2}(b) is linear, it does not carry any spin-angular momentum, so the observed rotations are only due to the radiation torque exerted by the light on the particles because their asymmetric shape. 

In the case of circular polarised light, depending on the absorption properties of the particle, a laser beam can induce a spin angular torque in addition to the radiation one  \citep{marston1984radiation,jones2015optical}. When a laser beam is circularly polarized, each of its photons carries a spin angular momentum  $+\hbar$ for left circularly polarised light (LCP) $-\hbar$ for right circularly polarised light (RCP). Therefore, the total torque acting on a non sperical particle is given by different contributions that include a radiation torque related to the shape and the transferred spin angular momentum. The spin angular momentum is added to to the "windmill effect" in the case of LCP and subtracted in the case of RCP \citep{jones2015optical}.

 %The latter case is the one of left circular polarised (LCP) and right circular polarised (RCP) light. In the case of  polarised light beams, we also observed rotations of the trapped particles, as showed in Fig. \ref{fig4}(a,b), where rotations are due now to the sum of two contributes: the radiation torque and the transfer of spin-angular momentum. 

 In Fig. \ref{fig2}(c) we report the rotational frequencies $\Omega_{\rm rot}$ of an optically trapped grain of the terrestrial samples HE-1 from Etna volcano for different light polarisation. Here, $\Omega_{\rm rot}$ decreases from  -18 $\rm s^{-1}$ for LCP to -28 $\rm s^{-1}$ for RCP with a central value of -23 $\rm s^{-1}$ for LP indicating that the trapped particle is absorbing circular polarised light with a spin angular momentum  $+\hbar$ for LCP light and $-\hbar$ for RCP light. Thus, negative frequency values mean a rotation anticlockwise with respect to the $z$ axis, while positive frequency values mean a rotation clockwise with respect to the $z$ axis. The rotational frequency $\Omega_{\rm rot}$ reported in Fig. \ref{fig2}(c) is the result of rotations induced by the total torque acting on the sample. The radiation pressure torque, which does not depend on light polarisation, produces a particle rotation frequency of -23 $\rm s^{-1}$, while the spin angular momentum torque sums or subtracts depending on the helicity of light. In the case of LCP light the contribution of the spin angular momentum torque is positive and summed to the radiation torque increasing the rotational frequency of the particle to -18 $\rm s^{-1}$, while in the case of RCP light the contribution is negative decreasing the rotational frequency of the particle to -28 $\rm s^{-1}$. This demonstrates that the total optomechanical interaction of particles with light, depends both on the particle shape and on the helicity of the laser beam.

\subsection{Spectroscopic analysis and mineralogy of individual grains}
We characterize the mineralogical compositions of terrestrial and extraterrestrial dust grains with different shapes and sizes by Raman tweezers, Fig. \ref{fig1}(d).
All the investigated sample are trapped in 3D or in 2D, whenever the radiation pressure is stronger than the axial gradient force. In the latter case, a stable 3D optical trapping is not achievable because the radiation pressure pushes the particle away along the $z$ direction, along which the laser beam propagates. To overcome this issue, we push the particle against the wall of the glass slide, thereby physically confining the particle along the $z$ direction, while the particle is still confined along the $x-y$ direction by the optical gradient forces, defining the optical potential. %In the latter case, 2D trapping is achieved by pushing the particle against the cell wall by radiation pressure.
For each sample, a reference background spectrum, without any trapped particle, is also acquired and used to remove the spurious signal from the glass slide and the media (water, immersion oil). Thereafter, by a fitting routine, the peaks of each spectrum are identified and  compared with reference spectra for the identification of the constituent phases in the trapped samples.
We first analyze the standard samples M26, HE-1 and A-1 in order to match their mineral composition reported in literature \citep{fazio2014shock, orlando2008experimental, brunetto2014ion, cristofolini1987petrological}.

\begin{figure}
  \centering
  \resizebox{14 cm}{!}{%
  \includegraphics{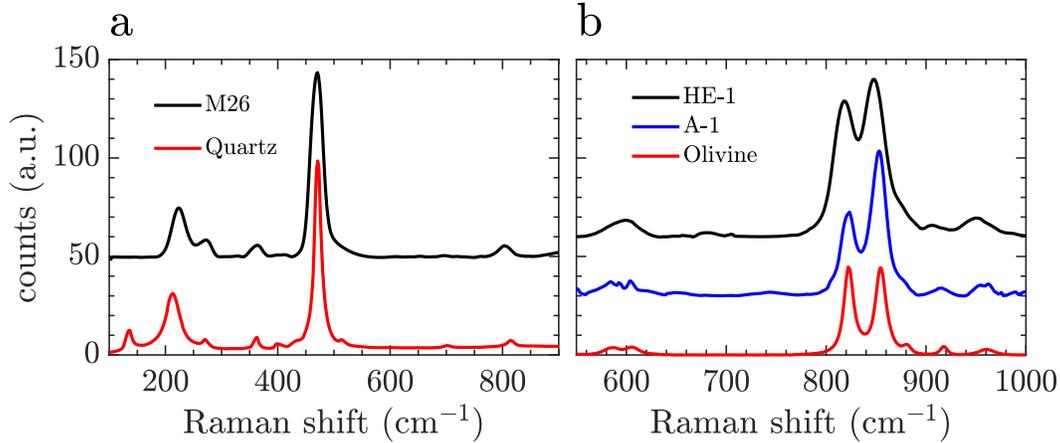}
  }
  \caption{Raman spectra of optically trapped single dust particle from sample: a) M26 from Kamil crater (black line), red line represents the Raman spectrum of quartz used as reference (RRUFF R150074). b) HE-1 from Etna (black line) and A-1 from Allende meteorite (blue line). Red line represents the Raman spectrum of olivine used as standard reference (RRUFF X050088). All spectra are offset for clarity.  }
  \label{fig4}
\end{figure}

In Fig. \ref{fig4}(a) we report a Raman spectrum  (black line) collected by a single optically trapped grain of sample M26. According to the literature it contains mainly quartz whose reference spectrum is represented by the red line \citep{fazio2014shock}. Similarly, in Fig. \ref{fig4}(b) we report the Raman spectra, of a single grain of the sample HE-1 from Etna volcano (black line) and of A-1 from Allende meteorite (blue line). According to the literature, both meteorites contain olivine, whose reference spectrum is represented by the red line \citep{cristofolini1987petrological}. For all the experimental spectra of our samples showed in Fig. \ref{fig4}(a,b) there is a good agreement with the reference spectra of their mineral composition found in literature.

\begin{figure}
  \centering
  \resizebox{14 cm}{!}{%
  \includegraphics{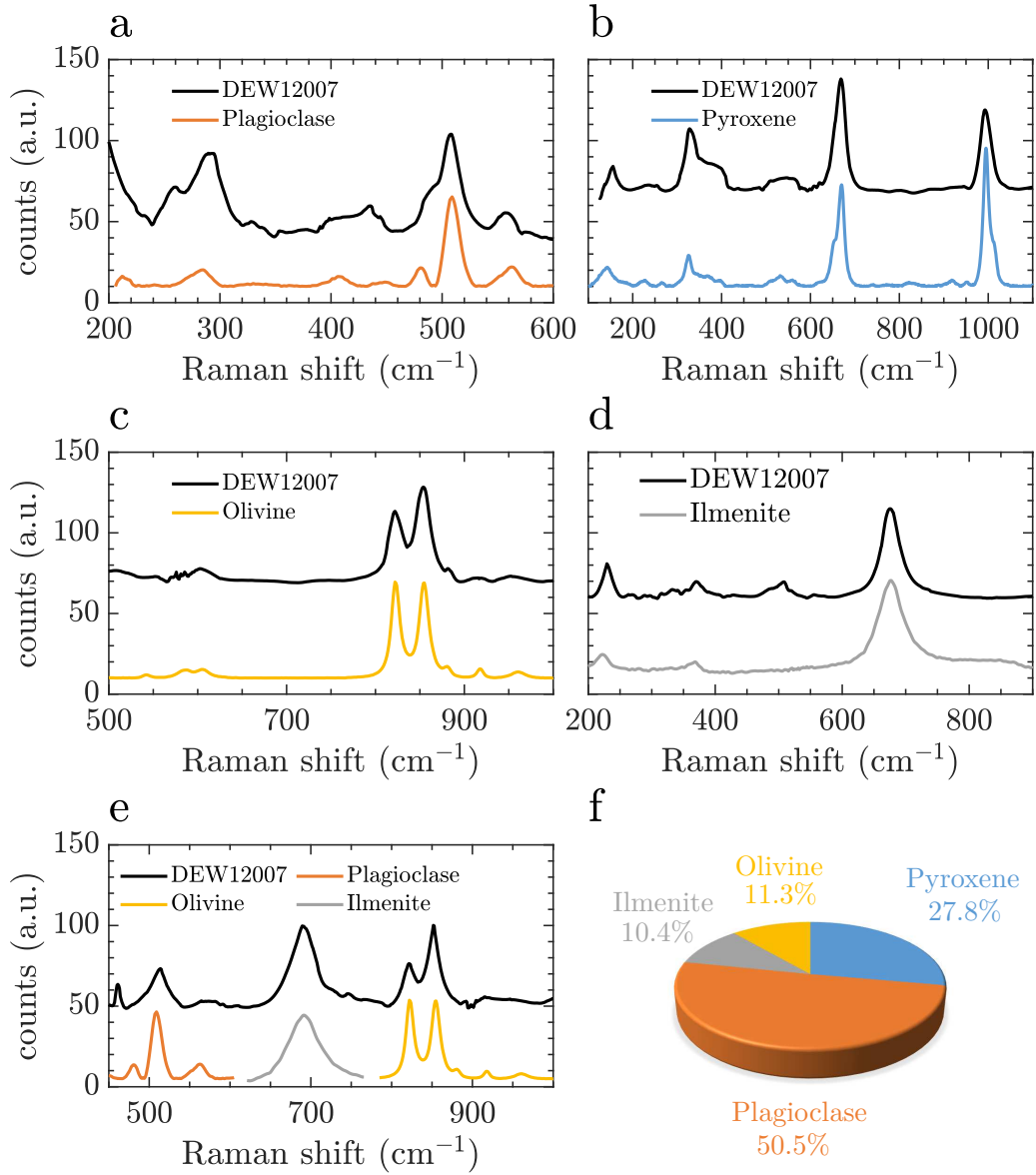}
  }
  \caption{Raman spectra and its mineralogical composition of optically trapped dust particles of the lunar meteorite DEW 12007. Black lines represent the Raman spectra of the trapped samples, while colored lines represent the reference Raman spectra from the RRUFF database of: a) plagioclase (RRUFF X050108), b) pyroxene (RRUFF R200002), c) olivine (RRUFF X050088) and d) ilmenite (RRUFF R060149), used for the mineral identification of the trapped single grains. All spectra are offset for clarity. e) Raman spectra of a single grain of DEW 12007 (black line) containing several mineral components (colored lines). f) Mineral occurrence of the mineral constituents over 70 trapped grains of the lunar meteorite DEW 12007. }
  \label{fig5}
\end{figure}

After testing our Raman tweezers on terrestrial and extraterrestrial standard  grains, we investigate cosmic dust from the lunar meteorite DEW 12007. In particular, we trap about 70 different particles in 2D and 3D, following an appropriate protocol to avoid to trap the same grains twice.
In Fig. \ref{fig5}(a-d) we report with black lines the Raman spectra of four different trapped grains of DEW 12007, and with colored lines the reference Raman spectra of some of their constituents according to the literature \citep{collareta2016high}. In particular, in Fig.  \ref{fig5}(a-d) we can observe that the trapped grains contain the following minerals: plagiocase (labradorite); pyroxene (augite and pigeonite); olivine (forsterite and fayalite); ilmenite.

Figures~\ref{fig4}(a,b) and \ref{fig5}(a,d) show that each single trapped grains contains only a single mineral component, among the several ones reported in literature, matching its own reference spectrum. We identified both monomineralic and polymineralic grains (Fig.~\ref{fig5}e), as expected for a rock in which the grain-size is highly variable.

The occurrence of the mineral constituents  of the lunar meteorite DEW 12007 over 70 dust grains is reported in Fig. \ref{fig5}(f), where it is possible to notice that plagioclase is the most recurring mineral constituent, with an occurrence of $50.4\%$, then we found that pyroxene is the second most abundant mineral with an occurrence of $27.8\%$, and finally olivine and ilmenite with an abundance of  $11.3\%$ and $10.4\%$ respectively.  The rank of the mineral abundances of the constituents of DEW 12007 matches the data from literature \citep{collareta2016high}. It is noteworthy that during the grinding process it could happen that cosmic dust was not homogeneously ground, producing grains with more than one or without any clear mineral component.

\section{Conclusions}

We used optical tweezers to trap individual micron-sized dust particles of astrophysical interest to characterize their opto-mechanical response. We calculated their trap stiffness and their rotational frequencies as a function of the trapping laser power, showing how these quantities increase almost linearly with the laser power. We also calculated the theoretical values of the optical forces acting on cosmic dust by T-matrix formalism, finding a good agreement between the experimental and theoretical values, validating our theoretical models. Moreover, we investigated the effect of light polarisation on cosmic dust, measuring the particle rotational frequency for different helicity of light, showing how the total opto-mechanical interaction depends on both the particle shape and the polarisation of the laser beam.
Furthermore, we used Raman tweezers to investigate the mineral composition of the dust particles. The agreement of our results with the literature validates optical trapping for the analysis of cosmic dust,  opening new perspectives in the investigation of extraterrestrial particles on our planet.   
Optical trapping techniques, thanks to their contactless and non-invasive unique combination of capabilities, can be used to maximize the scientific return from the
analyses of cosmic dust samples collected by the current and future sample return missions (e.g. OSIRIS-Rex, Mars 2020; Chang’e 5), particularly during the preliminary investigation procedures in receiving/curation facilities \citep{smith2021roadmap}. High-resolution, contactless and non-invasive analysis of planetary dust are actually expected to provide unprecedented information on the astrophysical origin and geologic evolution of their parent bodies. Furthermore, they are also expected to be instrumental for biohazard assessment for constrained sample return missions – like those targeting Mars or the icy bodies in the outer solar system, and the detection of past or extant extraterrestrial life. 
Finally, the successful trapping of particles in a water medium is the first step towards the realization of trapping in air/vacuum, a totally non-invasive micromanipulation fundamental for extraterrestrial materials. For these reasons, optical tweezers are key tools in controlled laboratory experiments, aiming for space  applications to trap and characterize dust particles directly in space or on extraterrestrial bodies during exploratory missions.

%can be implemented in extra-terrestrial curation facilities at our planet, where extra terrestrial samples are processed in order to prevent sample contamination and even more important to prevent bio-hazards for our planet.\\

\section{Acknowledgment}
We  acknowledge  financial contribution from the agreement ASI-INAF n.2018-16-HH.0, project “SPACE Tweezers” and from the  MSCA ITN project “ActiveMatter" sponsored by the European Commission (Horizon 2020, Project Number 812780). A. Musolino acknowledges support by ASI-INAF, “Rosetta GIADA”, I/024/12/0.
The lunar regolith breccia was collected during the 2017 Antarctic Campaign of the Programma Nazionale delle Ricerche in Antartide (PNRA) within the “Meteoriti Antartiche" project (ID: PNRA16\_00029) and the sample used in this study was provided by the Museo Nazionale dell'Antartide di Siena. Scanning electron microscopy (SEM) imaging was carried out the ESEM-FEG facility of the Centro per la Integrazione  della Strumentazione - Università di Pisa (CISUP).

\appendix

\section{Optical forces}
\label{AOF}

Forces arising in optical trapping are a consequence of the conservation of the electromagnetic momentum in the light-matter interaction  \citep{jones2015optical}.
However, some simplifications and approximations have been made for an easier understanding of optical forces. These approximate approaches are often a valuable source of physical insight and have a great pedagogical value  \citep{magazzu2015optical,polimeno2018optical, jones2015optical}.
In the geometrical optics approximation, valid for particle size larger than the wavelength of the trapping beam, optical forces can be divided in a gradient and a scattering component. The gradient force component is perpendicular to the propagation of the laser beam, it is proportional to the intensity gradient of the laser spot and it is responsible for trapping. While the scattering force component, having the same direction of the propagating laser beam, is   proportional to the light intensity and tends to push particles away from the laser focus due to the radiation pressure \citep{jones2015optical}.
In this approximation the incoming optical field can be considered as a collection of light rays, each of them carrying a portion of the total optical power and linear momentum. When a ray impinges on a particle, it is partly transmitted and partly reflected by the particle surface and the exchange of linear momentum between the ray and the particle generates an optical force $\vec F=\frac{\Delta \vec P}{\Delta t}$, where $\Delta \vec P=\vec P_{\rm inc}-\vec P_{\rm ref}$ is the exchanged momentum calculated as the difference between the incident and reflected momentum $\vec P_{\rm inc}$ and $\vec P_{\rm ref}$ respectively during a time interval $\Delta t$. The optical force exerted by a laser beam on a particle is the sum of the forces generated by each constituting ray and for specific experimental conditions and parameters (e.g. when the refractive index of a particle is higher than the one of its surrounding medium, proper size of the particle and laser power, etc.) these forces can confine a particle within a equilibrium region close to the focal spot, where the force acting on the particle is zero \citep{magazzu2015optical,polimeno2018optical, jones2015optical}.
When a trapped particle is displaced from the centre of the trap to a non equilibrium point, a net optical force acts like a restoring spring, bringing the particle back toward the centre of the trap. For small displacement of the particle from its equilibrium position the restoring force is, to a first approximation, proportional to the displacement acting like a Hookean spring with a fixed stiffness  \citep{jones2015optical, gieseler2021optical}, i.e 

\begin{equation}
F_x \approx - \kappa_{x}(x- x_{\rm eq}) 
 \label{hook}
  \end{equation}
where, considering for simplicity only a single component of the force, $x$ is the particle position, $x_{\rm eq}$ is the equilibrium position and $\kappa_{x}$ is the trap stiffness. The trap stiffness can be obtained by several calibration methods, \citep{jones2015optical, gieseler2021optical} and by its value we can quantify an external force $ F_{{\rm ext},x}$ acting on a trapped particle by measuring the particle displacement $\Delta x_{\rm eq}=(x- x_{\rm eq})$  from its equilibrium position, i.e. $F_{{\rm ext},x} = \kappa_{x}\Delta x_{\rm eq}$

\section{Experimental setups}
\label{AES}

The light source of the standard optical tweezers used for the investigation of cosmic dust is a laser diode (LD) generating a linear polarised laser beam with a wavelength of 830 nm. The laser beam is expanded by a two lenses ($\rm{l_1, l_2}$) telescope system and reflected by a dichroic mirror towards the back aperture of a high numerical aperture (NA) oil immersion objective, which was also used to imagine the sample on a CMOS camera, Fig. \ref{fig1}(e). Thanks to the telescope system, the beam overfills the back aperture of the objective giving rise to  a highly focused laser beam. The overfilling creates the maximal optical field gradient in the focal spot for a more efficient optical trapping. A sample holder is equipped with a 3D translation piezo-stage to move the focal spot within the cavity glass slide containing the sample solution. The dichroic mirror used to reflect the laser beam to the objective acts like a short pass filter, it reflects the laser light towards the objective and transmits the visible light to a CMOS camera, preventing the saturation of the detector and allowing a clear view of the sample on a monitor. The polarisation of the trapping beam can be tuned by a waveplate to investigate the optical response of cosmic dust for different light polarisations. In particular, changing the light polarisation from linear to right- or left-circular, the occurrence of spin-angular momentum transfer from light to the dust can be investigated \citep{donato2014polarization, jones2015optical}. 
The forward-scattered light from a trapped particle, containing information about the particle position, is collected together with the transmitted light by a condenser. The superposition of these two beams generates an interferometric pattern, which is reflected by a second dichroic mirror towards a quadrant photo diode (QPD) through a lens (l$_3$) Fig. \ref{fig1}(e) \citep{polimeno2018optical, jones2015optical}. A QPD converts the interferometric pattern collected by a condenser in analogical voltage signals, proportional to the displacement of the particle from its equilibrium position \citep{gittes1998interference, jones2015optical, magazzu2015optical}. The signals from the QPD were acquired at a sampling frequency of 50 kHz by a National Instrument  acquisition board for time intervals of 2 sec.  The sampled signals were then analysed by a PC %to rebuild the 3D trajectories of the trapped particle,
providing information about the opto-mechanical interaction between light and a single grain of cosmic dust. 

%calibration 
In order to characterize the optical trapping forces on a single dust grain, we extract some crucial parameters from the particle random trajectories in the confining potential, such as the trap relaxation frequency. The starting point of optical force measurements in optical tweezers is the description of the centre-of-mass random motion of an optically trapped particle through an overdamped Langevin equation \citep{jones2015optical}. In particular, for small displacements around the equilibrium point a positional coordinate, e.g. $x$, of the trapped particle, subject to the random force, $F_{\rm random}(t)$, can be expressed in terms of the following equation:

\begin{equation}
\gamma \frac{dx}{dt} +\kappa_xx=F_{\rm random}(t)
 \label{Lang}
  \end{equation}

where  $\gamma$ is the Stokes friction coefficient of a particle immersed in a fluid medium and $\kappa_x$ is the trap stiffness that we wish to measure. The first term of eq. \ref{Lang} is the velocity dependent viscous force, the second one is the restoring term due to the optical potential, and the final one describes the Brownian stochastic force. 
The power spectral density (PSD) of a signal describes how its energy is distributed in the frequency domain and it is useful to calibrate optical tweezers. Indeed, the power spectral density of eq. \ref{Lang} show a Lorentzian shape:

\begin{equation}
S_x(\omega)=\frac{2D}{(\omega^2+\omega_x^2)} 
 \label{PSD}
\end{equation}

where the half-width of the distribution is the relaxation frequency defined as $\omega_x=\kappa_x/\gamma$ and $D=k_BT/\gamma$ is the diffusion coefficient from the Stokes-Einstein equation \citep{jones2015optical}. By fitting the power spectrum of the particle coordinate with a Lorentzian shape function we obtain the relaxation frequency $\omega_x$ as a fit parameter and hence we can get the trap spring constant $\kappa_x$ \citep{gittes1998interference,  magazzu2015optical}.
%calibration 

The Raman tweezers used for the identification of the minerals constituents of our samples is a customised setup obtained by coupling a homemade optical tweezers  with a commercial Raman spectrometer (Horiba TRIAX 190) through a notch filter, Fig. \ref{fig1}(f). This filter reflects the laser light, from a laser diode having a wavelength of $780\  {\rm nm}$,  to the back aperture of an oil immersion objective, similarly to the case of the dichroic mirror previously described in the standard optical tweezers setup, Fig. \ref{fig1}(e).
%resulting in a maximum power of 7 mW at the output of the objective. 
The notch filter reflects only the single wavelength of the laser beam, and it is transparent to all the other wavelengths. In such a way, the elastic component of the scattering is cut out by the notch filter and only the Raman signal is transmitted to the spectrometer, which is equipped with a grating having a spectral resolution of $8\ \rm{cm^{-1}}$ and coupled to a silicon Peltier-cooled CCD camera to acquire the spectra. 
The Raman spectra of our investigated samples were obtained with a laser power of about $7\ {\rm mW}$ at the sample and acquired with an integration time of few tens of seconds.

\begin{figure}
  \centering
  \resizebox{14 cm}{!}{%
  \includegraphics{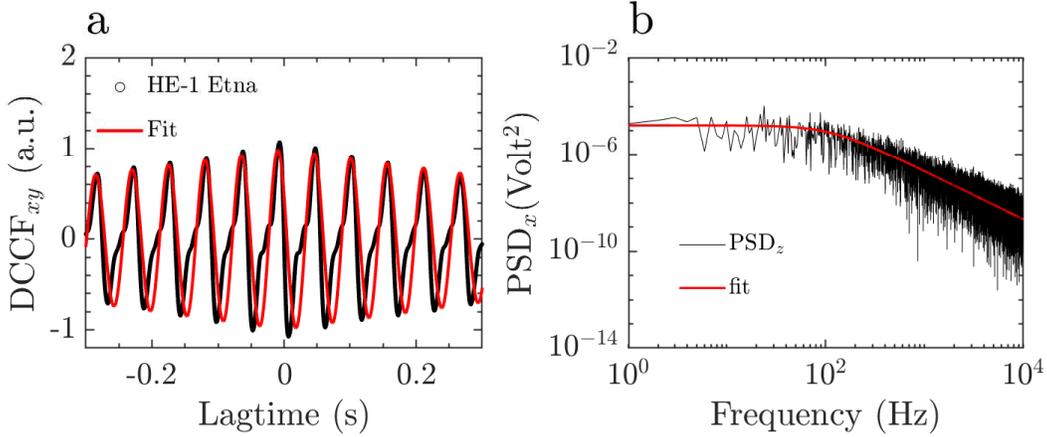}
  }
  \caption{Rotational and relaxation frequencies. a) Differential cross correlation  function ${\rm DCCF}_{xy}(\tau)$ of the sample HE-1 from Etna, black circles represent experimental data, while red line represents a sinusoidal fitting function. Discrepancy between the experimental data and the fit are due to the presence of of a second rotational or vibrational motion of the trapped particle due to its non-spherical geometry and anisotropy \citep{jones2009rotation}.    b) Black line represents the power spectrum density (PSD) of the trajectory along $z$ direction for a trapped dust grain of the lunar meteorite  DEW 12007. While the red line represent a Lorentzian fit of the calculated PSD, where the relaxation frequency of the trapped grain is obtained as fitted parameter.}
  \label{apx}
\end{figure}

\section{Theoretical calculations}
\label{ATH}

\subsection{Radiation force in optical tweezers}

In the framework of electromagnetic scattering theory, optical forces and optical trapping are the consequence of the electromagnetic momentum conservation during a light scattering process \citep{jones2015optical, borghese2007scattering}. Using the linear momentum conservation, the time-averaged optical force on a generic particle exerted by a monochromatic light is: 
\begin{equation}
\mathbf{F}_{\mathrm {rad}}=\oint_{S}\mathrm{\bar{T}_M}\cdot\hat{n}dS 
\end{equation}
where the integration is carried out over a surface S surrounding the scattering particle, $\hat{n}$ is the outward normal unit vector, $\vec{r}$ is the vector position, and $\mathrm{\bar{T}_M}$ is the averaged Maxwell stress tensor in the Minkowski form \citep{pfeifer2007}  describing the optomechanical interaction. For a non-magnetic medium \citep{borghese2007scattering}:

\begin{equation}
 \mathbf{F}_{\mathrm {rad}} = - \frac{\varepsilon_{\mathrm m} r^2}{4} \int_{\Omega} \left( |\textbf{E}_{\mathrm s}|^2 + \frac{c^2}{n_{\mathrm m}^2} |\textbf{B}_{\mathrm s}|^2 + 2\Re \left\{ \textbf{E}_{\mathrm i} \cdot\textbf{E}_{\mathrm s}^* + \frac{c^2}{n_\mathrm{m}^2}\textbf{B}_{\mathrm i} \cdot \textbf{B}_{\mathrm s}^* \right\} \right) \hat{\textbf{r}} \ \mathrm{d}\Omega
\end{equation}

\noindent where $\varepsilon_{\mathrm m}=\varepsilon_0 n_{\mathrm m}^2$ is the medium permettivity, $r$ and $\hat{r}$ are the modulus and the unit vector of the vector position $\vec{r}$ respectively, $\varepsilon_0$ is the vacuum permittivity, $n_{\mathrm m}$ is the medium refraction index, $c$ is the light velocity, $\mathbf{E}_{\mathrm i}$, $\mathbf{B}_{\mathrm i}$, $\mathbf{E}_{\mathrm s}$, $\mathbf{B}_{\mathrm s}$ are respectively the incident electric and magnetic field and the scattered electric and magnetic field and the integration is taken over the full solid angle $\Omega$. The scattered electromagnetic field $\mathbf{B}_{\mathrm s}$ is calculated in near zone, take into account the transmission properties of the wave inside the nanoparticle. When we deal with spherical monomers, the expression of the radiation force should be rewritten in terms of T-matrix formalism \citep{borghese2007scattering, saija2005transverse,polimeno2018optical}. Because of the linearity of Maxwell's equations, once the field involved in the scattering process are expanded in terms of Vector Helmholtz Harmonics (VHH), it is possible, through the definition of T-matrix, to obtain the relation between the incident and scattered field imposing the boundary conditions to the fields on the spherical surface \citep{borghese2007optical}. The T-matrix $\mathbb{T} \equiv \{ T^{(p'p)}_{l'm'lm} \}$ elements encompasses all the information on the morphology of the scattering particle with respect to the same incident field, binding the known multipole amplitudes of the incident field $W^{(p)}_{{\mathrm i}, lm}$ with the unknown amplitudes of the scattered field $A_{{\mathrm s},l'm'}^{(p')}$ \citep{borghese2007scattering}.
For the case of optical tweezers, we generalized the incident field resorting to the angular spectrum representation of Richards and Wolf \citep{richard1959} and then calculated the optical tweezers stiffness for each particle \citep{jones2015optical, borghese2007optical}.

\subsection{Optical properties of composite particles}

Knowledge of the optical properties of particles that interact with radiation is a key ingredient for the correct modelling of a matter radiation interaction. When the medium that constitutes the particles is a multi-component mixture, two main paths can be taken for modelling: (1) homogenization theory \citep{garnett1904,garnett1906} through the use of the Bruggeman mixing formula \citep{bruggeman1935,bruggeman1936}; (2) the representation of complexity by means of a random geometry of one-component inclusions whose size can respect the fraction of volume obtained from the experimental chemical-physical analysis of the sample.
The Bruggeman mixing formula \citep{bruggeman1935,bruggeman1936}  can be applied to a medium composed of N kinds of inclusions with permettivities $\varepsilon_n$ and volume fractions $f_n$ such that $\sum_{n} f_n =1$. In this case, unlike Maxwell Garnett Mixing Formula that is inapplicable when the volume fractions of all components are comparable, the Bruggeman mixing formula is symmetric with respect to all medium components and does not treat any one of them differently. Therefore, it can be applied, at least formally, to composites with arbitrary volume fractions without causing contradictions when we cannot distinguish the “host” material from the “inclusions”. From operative point of view, the Bruggeman effective permittivity, which we denote by $\varepsilon_{BG}$, can be obtained satisfying the following equation: 

\begin{equation}
\sum\limits_{n=1}^N f_n \frac{\varepsilon_{n} – \varepsilon_{BG}}{\varepsilon_n+2\varepsilon_{BG}}=0    
\end{equation}

\noindent where  $\sum_{n} f_n =1$.
In the case of a strongly inhomogeneous particle consisting of islands of different materials, the process of interaction underlying the optical trapping phenomena is reminiscent of the multi-component structure of the nanoparticle and the formula of Bruggeman, derived within the homogenization theory, tends to mask this behaviour. For this purpose,  the optical behaviours of the spherical inclusions contained within the spherical nanoparticle, which constitute an aggregate,  is studied by applying the theory of electromagnetic scattering to cluster of spheres. 

Within the T-matrix approach, the calculus of the multipole amplitudes of the field scattered by the
whole object is performed in two steps. First, we consider the superposition of the field that would exist within the sphere in absence of any inclusion and of the scattered fields coming from the aggregate  taking  into account all the multipolar interactions. This superposition constitutes the internal field of the host sphere. Second, we impose the boundary conditions to the
internal field and to the external field across the surface of the host sphere to get the amplitudes of the scattered field.

As a result of the first step, we get a system of linear nonhomogeneous equations that in symbolic form can be written as:

\begin{equation}
\mathbb{P}=\mathbb{Z}\mathbb{W}
\end{equation}
where $\mathbb{P}$ is the matrix whose elements are the unknown scattering coefficients of the internal field to be calculated, $\mathbb{Z}$ is the T-matrix of the internal aggregate, and $\mathbb{W}$  is the matrix of the known external incident field coefficients.

Once the amplitudes of $\mathbb{P}$ have been calculated, we proceed to the second step which in symbolic form  yields:

\begin{equation}
\mathbb{A}=\mathbb{S}\mathbb{P}=\mathbb{S}(\mathbb{Z}\mathbb{W})=\mathbb{T}\mathbb{W}
\end{equation}
where $\mathbb{A}$ is the matrix whose elements are the scattering coefficients in external medium to be calculated, S is the matrix which acts on the matrix $\mathbb{P}$ containing the coefficients of the already calculated internal field, and $\mathbb{T}$ is the transition matrix for the whole scatterer \citep{borghese1994eccentric,borghese1998eccentric}. 

\section{Samples characterization}
\label{ASC}

The samples used in this work (M-26, HE-1, A-1, DEW 12007) have been analysed under the field emission scanning electron microscope with energy dispersive spectroscopy FEG-SEM FEI QUANTA 450 at the Center for Instrument Sharing of the University of Pisa (CISUP) for a textural characterization (shape, grain size). Their main mineralogic composition is determined using X-ray powder diffraction data acquired at the Dipartimento di Scienze della Terra of the Università di Pisa. Lithology and mineralogic composition of the samples are reported in Table~\ref{tab:samplescharacterization}. 

The characterization of M-26 refers to literature \citep{fazio2014shock}. In-depth characterization of DEW 12007 can be found in literature\citep{collareta2016high}. Quantitative mineralogical compositions of the samples M-26 (quartzarenite), HE-1 (hawaiite), A-1 (carbonaceous chondrite Allende), and DEW 12007 (lunar regolith breccia) were obtained applying the Rietveld method \citep{bish1993quantitative} and given in Table~\ref{tab:samplescharacterization}. X-ray powder diffraction data were collected at the Dipartimento di Scienze della Terra of the Università di Pisa, on a Bruker D2 Phaser diffractometer, equipped with a Lynxeye detector, operating at 30 kV and 10 mA and using Cu K$\alpha$ radiation ($\lambda$ = 1.54184 Å). The diffraction patterns were collected over the 5-70$^\circ$ 2$\theta$ range for DEW 12007 and 12-70$^\circ$ 2$\theta$ range for HE-1 and Allende, with 0.02$^\circ$ scan step size and 1 $\rm s$ counting time per step. Samples were spiked with a known amount of rutile as internal standard (NIST SRM 674a) in order to check for the presence of an amorphous component \citep{gualtieri2000accuracy}.

\begin{table}[]
\caption{Textural and mineralogic composition of the samples used to test the Raman tweezers set-up: (1) M-26, quartzarenite; (2) HE-1 hawaiite; (3) A-1 Allende; (4) DEW 12007. For each sample is reported: (1) its lithology (type of rock and textural information) obtained at the FEG-SEM and literature and (2) the overall mineralogical composition [weight \%] obtained with X-ray diffraction.}
\label{tab:samplescharacterization}
\resizebox{\columnwidth}{!}{%
\begin{tabular}{lll}
\hline
\textbf{Sample}    & \begin{tabular}[c]{@{}l@{}}\textbf{Lithology}\end{tabular}                                                                                                                                                                                       & \begin{tabular}[c]{@{}l@{}}\textbf{Overall  min. compo. wt\%}\end{tabular}                                                                                                                                 \\ \hline
M-26               & \begin{tabular}[c]{@{}l@{}}Quarzarenite -- Coarse-grained  \\ \end{tabular}                                                                                                                                 & \begin{tabular}[c]{@{}l@{}}Quartz \textgreater{}99 \end{tabular}                                                                                                                                                                                          \\ \hline
HE-1               & \begin{tabular}[c]{@{}l@{}}Hawaiite – Fine-grained. Basalt with pyroxene \\ and plagioclase microphenocrystals with abun-\\ dant glass inclusions. Vesicular. 30\% phenocrysts\\  by volume\end{tabular} & \begin{tabular}[c]{@{}l@{}}Plagioclase: labradorite 56 \\ Clinopyroxene: augite 35\\ Olivine: forsterite 8\\ Magnetite \textless{}1\\ Glass \textless{}5\end{tabular}                                          \\ \hline
Allende (A-1)      & \begin{tabular}[c]{@{}l@{}}CV3 OxA   carbonaceous chondrite – chondrules\\  and CAIs ($\sim$55 vol\%) embedded in a fine-grained\\  dark matrix ($\sim$40 vol\%)\end{tabular}                            & \begin{tabular}[c]{@{}l@{}}Olivine: forsterite 80\\ Clinopyroxene: clinoenstatite 9\\ Clinopyroxene: diopside 7\\ Orthopyroxene: enstatite 2\\ Clinopyroxene: pigeonite 1\\ Chromite \textless{}1\end{tabular} \\ \hline
DeWitt (DEW) 12007 & \begin{tabular}[c]{@{}l@{}}Lunar polymictic breccia  with   \textless{}1 mm sized lithic\\  clasts in aphanitic matrix\end{tabular}                                                                      & \begin{tabular}[c]{@{}l@{}}Plagioclase: labradorite 48\\ Clinopyroxene: augite 23\\ Clinopyroxene: pigeonite 18\\ Olivine: forsterite 10\\ Ilmenite 1\\ Chromite \textless{}1\end{tabular}                     \\ \hline
\end{tabular}%
}
\end{table}

\section{Note}

A. Magazz\`u, D. Bronte Ciriza, and  A. Musolino contributed equally to this work.

\bibliography{Manuscript}{}

\begin{thebibliography}{66}%
\makeatletter
\providecommand \@ifxundefined [1]{%
 \@ifx{#1\undefined}
}%
\providecommand \@ifnum [1]{%
 \ifnum #1\expandafter \@firstoftwo
 \else \expandafter \@secondoftwo
 \fi
}%
\providecommand \@ifx [1]{%
 \ifx #1\expandafter \@firstoftwo
 \else \expandafter \@secondoftwo
 \fi
}%
\providecommand \natexlab [1]{#1}%
\providecommand \enquote  [1]{``#1''}%
\providecommand \bibnamefont  [1]{#1}%
\providecommand \bibfnamefont [1]{#1}%
\providecommand \citenamefont [1]{#1}%
\providecommand \href@noop [0]{\@secondoftwo}%
\providecommand \href [0]{\begingroup \@sanitize@url \@href}%
\providecommand \@href[1]{\@@startlink{#1}\@@href}%
\providecommand \@@href[1]{\endgroup#1\@@endlink}%
\providecommand \@sanitize@url [0]{\catcode `\\12\catcode `\$12\catcode
  `\&12\catcode `\#12\catcode `\^12\catcode `\_12\catcode `\%12\relax}%
\providecommand \@@startlink[1]{}%
\providecommand \@@endlink[0]{}%
\providecommand \url  [0]{\begingroup\@sanitize@url \@url }%
\providecommand \@url [1]{\endgroup\@href {#1}{\urlprefix }}%
\providecommand \urlprefix  [0]{URL }%
\providecommand \Eprint [0]{\href }%
\providecommand \doibase [0]{http://dx.doi.org/}%
\providecommand \selectlanguage [0]{\@gobble}%
\providecommand \bibinfo  [0]{\@secondoftwo}%
\providecommand \bibfield  [0]{\@secondoftwo}%
\providecommand \translation [1]{[#1]}%
\providecommand \BibitemOpen [0]{}%
\providecommand \bibitemStop [0]{}%
\providecommand \bibitemNoStop [0]{.\EOS\space}%
\providecommand \EOS [0]{\spacefactor3000\relax}%
\providecommand \BibitemShut  [1]{\csname bibitem#1\endcsname}%
\let\auto@bib@innerbib\@empty
%</preamble>
\bibitem [{\citenamefont {Calura}, \citenamefont {Pipino},\ and\ \citenamefont
  {Matteucci}(2008)}]{calura2008cycle}%
  \BibitemOpen
  \bibfield  {author} {\bibinfo {author} {\bibfnamefont {F.}~\bibnamefont
  {Calura}}, \bibinfo {author} {\bibfnamefont {A.}~\bibnamefont {Pipino}}, \
  and\ \bibinfo {author} {\bibfnamefont {F.}~\bibnamefont {Matteucci}},\
  }\bibfield  {title} {\enquote {\bibinfo {title} {The cycle of interstellar
  dust in galaxies of different morphological types},}\ }\href@noop {}
  {\bibfield  {journal} {\bibinfo  {journal} {Astronomy \& Astrophysics}\
  }\textbf {\bibinfo {volume} {479}},\ \bibinfo {pages} {669--685} (\bibinfo
  {year} {2008})}\BibitemShut {NoStop}%
\bibitem [{\citenamefont {Woosley}, \citenamefont {Heger},\ and\ \citenamefont
  {Weaver}(2002)}]{woosley2002evolution}%
  \BibitemOpen
  \bibfield  {author} {\bibinfo {author} {\bibfnamefont {S.~E.}\ \bibnamefont
  {Woosley}}, \bibinfo {author} {\bibfnamefont {A.}~\bibnamefont {Heger}}, \
  and\ \bibinfo {author} {\bibfnamefont {T.~A.}\ \bibnamefont {Weaver}},\
  }\bibfield  {title} {\enquote {\bibinfo {title} {The evolution and explosion
  of massive stars},}\ }\href@noop {} {\bibfield  {journal} {\bibinfo
  {journal} {Reviews of modern physics}\ }\textbf {\bibinfo {volume} {74}},\
  \bibinfo {pages} {1015} (\bibinfo {year} {2002})}\BibitemShut {NoStop}%
\bibitem [{\citenamefont {Woosley}\ and\ \citenamefont
  {Weaver}(1995)}]{woosley1995evolution}%
  \BibitemOpen
  \bibfield  {author} {\bibinfo {author} {\bibfnamefont {S.}~\bibnamefont
  {Woosley}}\ and\ \bibinfo {author} {\bibfnamefont {T.~A.}\ \bibnamefont
  {Weaver}},\ }\href@noop {} {\enquote {\bibinfo {title} {The evolution and
  explosion of massive stars ii: Explosive hydrodynamics and
  nucleosynthesis},}\ }\bibinfo {type} {Tech. Rep.}\ (\bibinfo  {institution}
  {Lawrence Livermore National Lab., CA (United States)},\ \bibinfo {year}
  {1995})\BibitemShut {NoStop}%
\bibitem [{\citenamefont {Rietmeijer}(1998)}]{rietmeijer1998interplanetary}%
  \BibitemOpen
  \bibfield  {author} {\bibinfo {author} {\bibfnamefont {F.~J.}\ \bibnamefont
  {Rietmeijer}},\ }\bibfield  {title} {\enquote {\bibinfo {title}
  {Interplanetary dust particles},}\ }in\ \href@noop {} {\emph {\bibinfo
  {booktitle} {Planetary Materials}}},\ Vol.~\bibinfo {volume} {36}\ (\bibinfo
  {publisher} {De Gruyter},\ \bibinfo {year} {1998})\BibitemShut {NoStop}%
\bibitem [{\citenamefont {Draine}(2003)}]{draine2003interstellar}%
  \BibitemOpen
  \bibfield  {author} {\bibinfo {author} {\bibfnamefont {B.~T.}\ \bibnamefont
  {Draine}},\ }\bibfield  {title} {\enquote {\bibinfo {title} {Interstellar
  dust grains},}\ }\href@noop {} {\bibfield  {journal} {\bibinfo  {journal}
  {Annual Review of Astronomy and Astrophysics}\ }\textbf {\bibinfo {volume}
  {41}},\ \bibinfo {pages} {241--289} (\bibinfo {year} {2003})}\BibitemShut
  {NoStop}%
\bibitem [{\citenamefont {Lodders}\ and\ \citenamefont
  {Amari}(2005)}]{loddersamari2005presolar}%
  \BibitemOpen
  \bibfield  {author} {\bibinfo {author} {\bibfnamefont {K.}~\bibnamefont
  {Lodders}}\ and\ \bibinfo {author} {\bibfnamefont {S.}~\bibnamefont
  {Amari}},\ }\bibfield  {title} {\enquote {\bibinfo {title} {Presolar grains
  from meteorites: Remnants from the early times of the solar system},}\
  }\href@noop {} {\bibfield  {journal} {\bibinfo  {journal} {Geochemistry}\
  }\textbf {\bibinfo {volume} {65}},\ \bibinfo {pages} {93--166} (\bibinfo
  {year} {2005})}\BibitemShut {NoStop}%
\bibitem [{\citenamefont {Westphal}\ \emph {et~al.}(2014)\citenamefont
  {Westphal}, \citenamefont {Stroud}, \citenamefont {Bechtel}, \citenamefont
  {Brenker}, \citenamefont {Butterworth}, \citenamefont {Flynn}, \citenamefont
  {Frank}, \citenamefont {Gainsforth}, \citenamefont {Hillier}, \citenamefont
  {Postberg} \emph {et~al.}}]{westphal2014evidence}%
  \BibitemOpen
  \bibfield  {author} {\bibinfo {author} {\bibfnamefont {A.~J.}\ \bibnamefont
  {Westphal}}, \bibinfo {author} {\bibfnamefont {R.~M.}\ \bibnamefont
  {Stroud}}, \bibinfo {author} {\bibfnamefont {H.~A.}\ \bibnamefont {Bechtel}},
  \bibinfo {author} {\bibfnamefont {F.~E.}\ \bibnamefont {Brenker}}, \bibinfo
  {author} {\bibfnamefont {A.~L.}\ \bibnamefont {Butterworth}}, \bibinfo
  {author} {\bibfnamefont {G.~J.}\ \bibnamefont {Flynn}}, \bibinfo {author}
  {\bibfnamefont {D.~R.}\ \bibnamefont {Frank}}, \bibinfo {author}
  {\bibfnamefont {Z.}~\bibnamefont {Gainsforth}}, \bibinfo {author}
  {\bibfnamefont {J.~K.}\ \bibnamefont {Hillier}}, \bibinfo {author}
  {\bibfnamefont {F.}~\bibnamefont {Postberg}},  \emph {et~al.},\ }\bibfield
  {title} {\enquote {\bibinfo {title} {Evidence for interstellar origin of
  seven dust particles collected by the stardust spacecraft},}\ }\href@noop {}
  {\bibfield  {journal} {\bibinfo  {journal} {science}\ }\textbf {\bibinfo
  {volume} {345}},\ \bibinfo {pages} {786--791} (\bibinfo {year}
  {2014})}\BibitemShut {NoStop}%
\bibitem [{\citenamefont {Brownlee}\ \emph {et~al.}(2003)\citenamefont
  {Brownlee}, \citenamefont {Tsou}, \citenamefont {Anderson}, \citenamefont
  {Hanner}, \citenamefont {Newburn}, \citenamefont {Sekanina}, \citenamefont
  {Clark}, \citenamefont {H{\"o}rz}, \citenamefont {Zolensky}, \citenamefont
  {Kissel} \emph {et~al.}}]{brownlee2003stardust}%
  \BibitemOpen
  \bibfield  {author} {\bibinfo {author} {\bibfnamefont {D.}~\bibnamefont
  {Brownlee}}, \bibinfo {author} {\bibfnamefont {P.}~\bibnamefont {Tsou}},
  \bibinfo {author} {\bibfnamefont {J.}~\bibnamefont {Anderson}}, \bibinfo
  {author} {\bibfnamefont {M.}~\bibnamefont {Hanner}}, \bibinfo {author}
  {\bibfnamefont {R.}~\bibnamefont {Newburn}}, \bibinfo {author} {\bibfnamefont
  {Z.}~\bibnamefont {Sekanina}}, \bibinfo {author} {\bibfnamefont
  {B.}~\bibnamefont {Clark}}, \bibinfo {author} {\bibfnamefont
  {F.}~\bibnamefont {H{\"o}rz}}, \bibinfo {author} {\bibfnamefont
  {M.}~\bibnamefont {Zolensky}}, \bibinfo {author} {\bibfnamefont
  {J.}~\bibnamefont {Kissel}},  \emph {et~al.},\ }\bibfield  {title} {\enquote
  {\bibinfo {title} {Stardust: Comet and interstellar dust sample return
  mission},}\ }\href@noop {} {\bibfield  {journal} {\bibinfo  {journal}
  {Journal of Geophysical Research: Planets}\ }\textbf {\bibinfo {volume}
  {108}} (\bibinfo {year} {2003})}\BibitemShut {NoStop}%
\bibitem [{\citenamefont {Frank}\ \emph {et~al.}(2014)\citenamefont {Frank},
  \citenamefont {Westphal}, \citenamefont {Zolensky}, \citenamefont
  {Gainsforth}, \citenamefont {Butterworth}, \citenamefont {Bastien},
  \citenamefont {Allen}, \citenamefont {Anderson}, \citenamefont {Ansari},
  \citenamefont {Bajt} \emph {et~al.}}]{frank2014stardust}%
  \BibitemOpen
  \bibfield  {author} {\bibinfo {author} {\bibfnamefont {D.~R.}\ \bibnamefont
  {Frank}}, \bibinfo {author} {\bibfnamefont {A.~J.}\ \bibnamefont {Westphal}},
  \bibinfo {author} {\bibfnamefont {M.~E.}\ \bibnamefont {Zolensky}}, \bibinfo
  {author} {\bibfnamefont {Z.}~\bibnamefont {Gainsforth}}, \bibinfo {author}
  {\bibfnamefont {A.~L.}\ \bibnamefont {Butterworth}}, \bibinfo {author}
  {\bibfnamefont {R.~K.}\ \bibnamefont {Bastien}}, \bibinfo {author}
  {\bibfnamefont {C.}~\bibnamefont {Allen}}, \bibinfo {author} {\bibfnamefont
  {D.}~\bibnamefont {Anderson}}, \bibinfo {author} {\bibfnamefont
  {A.}~\bibnamefont {Ansari}}, \bibinfo {author} {\bibfnamefont
  {S.}~\bibnamefont {Bajt}},  \emph {et~al.},\ }\bibfield  {title} {\enquote
  {\bibinfo {title} {Stardust interstellar preliminary examination ii: Curating
  the interstellar dust collector, picokeystones, and sources of impact
  tracks},}\ }\href@noop {} {\bibfield  {journal} {\bibinfo  {journal}
  {Meteoritics \& Planetary Science}\ }\textbf {\bibinfo {volume} {49}},\
  \bibinfo {pages} {1522--1547} (\bibinfo {year} {2014})}\BibitemShut {NoStop}%
\bibitem [{\citenamefont {Testa~Jr}\ \emph {et~al.}(1990)\citenamefont
  {Testa~Jr}, \citenamefont {Stephens}, \citenamefont {Berg}, \citenamefont
  {Cahill}, \citenamefont {Onaka}, \citenamefont {Nakada}, \citenamefont
  {Arnold}, \citenamefont {Fong},\ and\ \citenamefont
  {Sperry}}]{testa1990collection}%
  \BibitemOpen
  \bibfield  {author} {\bibinfo {author} {\bibfnamefont {J.~P.}\ \bibnamefont
  {Testa~Jr}}, \bibinfo {author} {\bibfnamefont {J.~R.}\ \bibnamefont
  {Stephens}}, \bibinfo {author} {\bibfnamefont {W.~W.}\ \bibnamefont {Berg}},
  \bibinfo {author} {\bibfnamefont {T.~A.}\ \bibnamefont {Cahill}}, \bibinfo
  {author} {\bibfnamefont {T.}~\bibnamefont {Onaka}}, \bibinfo {author}
  {\bibfnamefont {Y.}~\bibnamefont {Nakada}}, \bibinfo {author} {\bibfnamefont
  {J.~R.}\ \bibnamefont {Arnold}}, \bibinfo {author} {\bibfnamefont
  {N.}~\bibnamefont {Fong}}, \ and\ \bibinfo {author} {\bibfnamefont {P.~D.}\
  \bibnamefont {Sperry}},\ }\bibfield  {title} {\enquote {\bibinfo {title}
  {Collection of microparticles at high balloon altitudes in the
  stratosphere},}\ }\href@noop {} {\bibfield  {journal} {\bibinfo  {journal}
  {Earth and planetary science letters}\ }\textbf {\bibinfo {volume} {98}},\
  \bibinfo {pages} {287--302} (\bibinfo {year} {1990})}\BibitemShut {NoStop}%
\bibitem [{\citenamefont {Taylor}, \citenamefont {Baggaley},\ and\
  \citenamefont {Steel}(1996)}]{taylor1996discovery}%
  \BibitemOpen
  \bibfield  {author} {\bibinfo {author} {\bibfnamefont {A.}~\bibnamefont
  {Taylor}}, \bibinfo {author} {\bibfnamefont {W.}~\bibnamefont {Baggaley}}, \
  and\ \bibinfo {author} {\bibfnamefont {D.}~\bibnamefont {Steel}},\ }\bibfield
   {title} {\enquote {\bibinfo {title} {Discovery of interstellar dust entering
  the earth's atmosphere},}\ }\href@noop {} {\bibfield  {journal} {\bibinfo
  {journal} {Nature}\ }\textbf {\bibinfo {volume} {380}},\ \bibinfo {pages}
  {323--325} (\bibinfo {year} {1996})}\BibitemShut {NoStop}%
\bibitem [{\citenamefont {Della~Corte}\ \emph {et~al.}(2012)\citenamefont
  {Della~Corte}, \citenamefont {Palumbo}, \citenamefont {Rotundi},
  \citenamefont {De~Angelis}, \citenamefont {Rietmeijer}, \citenamefont
  {Bussoletti}, \citenamefont {Ciucci}, \citenamefont {Ferrari}, \citenamefont
  {Galluzzi},\ and\ \citenamefont {Zona}}]{della2012insitu}%
  \BibitemOpen
  \bibfield  {author} {\bibinfo {author} {\bibfnamefont {V.}~\bibnamefont
  {Della~Corte}}, \bibinfo {author} {\bibfnamefont {P.}~\bibnamefont
  {Palumbo}}, \bibinfo {author} {\bibfnamefont {A.}~\bibnamefont {Rotundi}},
  \bibinfo {author} {\bibfnamefont {S.}~\bibnamefont {De~Angelis}}, \bibinfo
  {author} {\bibfnamefont {F.~J.}\ \bibnamefont {Rietmeijer}}, \bibinfo
  {author} {\bibfnamefont {E.}~\bibnamefont {Bussoletti}}, \bibinfo {author}
  {\bibfnamefont {A.}~\bibnamefont {Ciucci}}, \bibinfo {author} {\bibfnamefont
  {M.}~\bibnamefont {Ferrari}}, \bibinfo {author} {\bibfnamefont
  {V.}~\bibnamefont {Galluzzi}}, \ and\ \bibinfo {author} {\bibfnamefont
  {E.}~\bibnamefont {Zona}},\ }\bibfield  {title} {\enquote {\bibinfo {title}
  {In situ collection of refractory dust in the upper stratosphere: The duster
  facility},}\ }\href@noop {} {\bibfield  {journal} {\bibinfo  {journal} {Space
  Science Reviews}\ }\textbf {\bibinfo {volume} {169}},\ \bibinfo {pages}
  {159–180} (\bibinfo {year} {2012})}\BibitemShut {NoStop}%
\bibitem [{\citenamefont {Lauretta}\ \emph {et~al.}(2017)\citenamefont
  {Lauretta}, \citenamefont {Balram-Knutson}, \citenamefont {Beshore},
  \citenamefont {Boynton}, \citenamefont {d’Aubigny}, \citenamefont
  {DellaGiustina}, \citenamefont {Enos}, \citenamefont {Golish}, \citenamefont
  {Hergenrother}, \citenamefont {Howell} \emph {et~al.}}]{lauretta2017osiris}%
  \BibitemOpen
  \bibfield  {author} {\bibinfo {author} {\bibfnamefont {D.}~\bibnamefont
  {Lauretta}}, \bibinfo {author} {\bibfnamefont {S.}~\bibnamefont
  {Balram-Knutson}}, \bibinfo {author} {\bibfnamefont {E.}~\bibnamefont
  {Beshore}}, \bibinfo {author} {\bibfnamefont {W.}~\bibnamefont {Boynton}},
  \bibinfo {author} {\bibfnamefont {C.~D.}\ \bibnamefont {d’Aubigny}},
  \bibinfo {author} {\bibfnamefont {D.}~\bibnamefont {DellaGiustina}}, \bibinfo
  {author} {\bibfnamefont {H.}~\bibnamefont {Enos}}, \bibinfo {author}
  {\bibfnamefont {D.}~\bibnamefont {Golish}}, \bibinfo {author} {\bibfnamefont
  {C.}~\bibnamefont {Hergenrother}}, \bibinfo {author} {\bibfnamefont
  {E.}~\bibnamefont {Howell}},  \emph {et~al.},\ }\bibfield  {title} {\enquote
  {\bibinfo {title} {Osiris-rex: sample return from asteroid (101955) bennu},}\
  }\href@noop {} {\bibfield  {journal} {\bibinfo  {journal} {Space Science
  Reviews}\ }\textbf {\bibinfo {volume} {212}},\ \bibinfo {pages} {925--984}
  (\bibinfo {year} {2017})}\BibitemShut {NoStop}%
\bibitem [{\citenamefont {Genge}\ \emph {et~al.}(2008)\citenamefont {Genge},
  \citenamefont {Engrand}, \citenamefont {Gounelle},\ and\ \citenamefont
  {Taylor}}]{genge2008classification}%
  \BibitemOpen
  \bibfield  {author} {\bibinfo {author} {\bibfnamefont {M.~J.}\ \bibnamefont
  {Genge}}, \bibinfo {author} {\bibfnamefont {C.}~\bibnamefont {Engrand}},
  \bibinfo {author} {\bibfnamefont {M.}~\bibnamefont {Gounelle}}, \ and\
  \bibinfo {author} {\bibfnamefont {S.}~\bibnamefont {Taylor}},\ }\bibfield
  {title} {\enquote {\bibinfo {title} {The classification of
  micrometeorites},}\ }\href@noop {} {\bibfield  {journal} {\bibinfo  {journal}
  {Meteoritics \& Planetary Science}\ }\textbf {\bibinfo {volume} {43}},\
  \bibinfo {pages} {497--515} (\bibinfo {year} {2008})}\BibitemShut {NoStop}%
\bibitem [{\citenamefont {Folco}\ and\ \citenamefont
  {Cordier}(2015)}]{folco2015micrometeorites}%
  \BibitemOpen
  \bibfield  {author} {\bibinfo {author} {\bibfnamefont {L.}~\bibnamefont
  {Folco}}\ and\ \bibinfo {author} {\bibfnamefont {C.}~\bibnamefont
  {Cordier}},\ }\bibfield  {title} {\enquote {\bibinfo {title}
  {Micrometeorites},}\ }\href@noop {} {\  (\bibinfo {year} {2015})}\BibitemShut
  {NoStop}%
\bibitem [{\citenamefont {Taylor}, \citenamefont {Messenger},\ and\
  \citenamefont {Folco}(2016)}]{taylor2016cosmic}%
  \BibitemOpen
  \bibfield  {author} {\bibinfo {author} {\bibfnamefont {S.}~\bibnamefont
  {Taylor}}, \bibinfo {author} {\bibfnamefont {S.}~\bibnamefont {Messenger}}, \
  and\ \bibinfo {author} {\bibfnamefont {L.}~\bibnamefont {Folco}},\ }\bibfield
   {title} {\enquote {\bibinfo {title} {Cosmic dust: finding a needle in a
  haystack},}\ }\href@noop {} {\bibfield  {journal} {\bibinfo  {journal}
  {Elements}\ }\textbf {\bibinfo {volume} {12}},\ \bibinfo {pages} {171--176}
  (\bibinfo {year} {2016})}\BibitemShut {NoStop}%
\bibitem [{\citenamefont {Rietmeijer}(2001)}]{rietmeijer2001identification}%
  \BibitemOpen
  \bibfield  {author} {\bibinfo {author} {\bibfnamefont {F.~J.}\ \bibnamefont
  {Rietmeijer}},\ }\bibfield  {title} {\enquote {\bibinfo {title}
  {Identification of fe-rich meteoric dust},}\ }\href@noop {} {\bibfield
  {journal} {\bibinfo  {journal} {Planetary and Space Science}\ }\textbf
  {\bibinfo {volume} {49}},\ \bibinfo {pages} {71--77} (\bibinfo {year}
  {2001})}\BibitemShut {NoStop}%
\bibitem [{\citenamefont {Rauf}\ \emph {et~al.}(2010)\citenamefont {Rauf},
  \citenamefont {Hann}, \citenamefont {Wallis},\ and\ \citenamefont
  {Wickramasinghe}}]{rauf2010evidence}%
  \BibitemOpen
  \bibfield  {author} {\bibinfo {author} {\bibfnamefont {K.}~\bibnamefont
  {Rauf}}, \bibinfo {author} {\bibfnamefont {A.}~\bibnamefont {Hann}}, \bibinfo
  {author} {\bibfnamefont {M.}~\bibnamefont {Wallis}}, \ and\ \bibinfo {author}
  {\bibfnamefont {N.}~\bibnamefont {Wickramasinghe}},\ }\bibfield  {title}
  {\enquote {\bibinfo {title} {Evidence for putative microfossils in space dust
  from the stratosphere},}\ }\href@noop {} {\bibfield  {journal} {\bibinfo
  {journal} {Int. J. Astrobiology}\ }\textbf {\bibinfo {volume} {9}},\ \bibinfo
  {pages} {183--189} (\bibinfo {year} {2010})}\BibitemShut {NoStop}%
\bibitem [{\citenamefont {Della~Corte}\ \emph {et~al.}(2014)\citenamefont
  {Della~Corte}, \citenamefont {Rietmeijer}, \citenamefont {Rotundi},\ and\
  \citenamefont {Ferrari}}]{della2014introducing}%
  \BibitemOpen
  \bibfield  {author} {\bibinfo {author} {\bibfnamefont {V.}~\bibnamefont
  {Della~Corte}}, \bibinfo {author} {\bibfnamefont {F.~J.}\ \bibnamefont
  {Rietmeijer}}, \bibinfo {author} {\bibfnamefont {A.}~\bibnamefont {Rotundi}},
  \ and\ \bibinfo {author} {\bibfnamefont {M.}~\bibnamefont {Ferrari}},\
  }\bibfield  {title} {\enquote {\bibinfo {title} {Introducing a new
  stratospheric dust-collecting system with potential use for upper atmospheric
  microbiology investigations},}\ }\href@noop {} {\bibfield  {journal}
  {\bibinfo  {journal} {Astrobiology}\ }\textbf {\bibinfo {volume} {14}},\
  \bibinfo {pages} {694--705} (\bibinfo {year} {2014})}\BibitemShut {NoStop}%
\bibitem [{\citenamefont {Mackinnon}\ and\ \citenamefont
  {Rietmeijer}(1987)}]{mackinnon1987mineralogy}%
  \BibitemOpen
  \bibfield  {author} {\bibinfo {author} {\bibfnamefont {I.~D.}\ \bibnamefont
  {Mackinnon}}\ and\ \bibinfo {author} {\bibfnamefont {F.~J.}\ \bibnamefont
  {Rietmeijer}},\ }\bibfield  {title} {\enquote {\bibinfo {title} {Mineralogy
  of chondritic interplanetary dust particles},}\ }\href@noop {} {\bibfield
  {journal} {\bibinfo  {journal} {Reviews of Geophysics}\ }\textbf {\bibinfo
  {volume} {25}},\ \bibinfo {pages} {1527--1553} (\bibinfo {year}
  {1987})}\BibitemShut {NoStop}%
\bibitem [{\citenamefont {Lewis}\ \emph {et~al.}(1987)\citenamefont {Lewis},
  \citenamefont {Ming}, \citenamefont {Wacker}, \citenamefont {Anders},\ and\
  \citenamefont {Steel}}]{lewis1987interstellar}%
  \BibitemOpen
  \bibfield  {author} {\bibinfo {author} {\bibfnamefont {R.~S.}\ \bibnamefont
  {Lewis}}, \bibinfo {author} {\bibfnamefont {T.}~\bibnamefont {Ming}},
  \bibinfo {author} {\bibfnamefont {J.~F.}\ \bibnamefont {Wacker}}, \bibinfo
  {author} {\bibfnamefont {E.}~\bibnamefont {Anders}}, \ and\ \bibinfo {author}
  {\bibfnamefont {E.}~\bibnamefont {Steel}},\ }\bibfield  {title} {\enquote
  {\bibinfo {title} {Interstellar diamonds in meteorites},}\ }\href@noop {}
  {\bibfield  {journal} {\bibinfo  {journal} {Nature}\ }\textbf {\bibinfo
  {volume} {326}},\ \bibinfo {pages} {160--162} (\bibinfo {year}
  {1987})}\BibitemShut {NoStop}%
\bibitem [{\citenamefont {Rotundi}\ \emph {et~al.}(2008)\citenamefont
  {Rotundi}, \citenamefont {Baratta}, \citenamefont {Borg}, \citenamefont
  {Brucato}, \citenamefont {Busemann}, \citenamefont {Colangeli}, \citenamefont
  {d'Hendecourt}, \citenamefont {Djouadi}, \citenamefont {Ferrini},
  \citenamefont {Franchi} \emph {et~al.}}]{rotundi2008combined}%
  \BibitemOpen
  \bibfield  {author} {\bibinfo {author} {\bibfnamefont {A.}~\bibnamefont
  {Rotundi}}, \bibinfo {author} {\bibfnamefont {G.}~\bibnamefont {Baratta}},
  \bibinfo {author} {\bibfnamefont {J.}~\bibnamefont {Borg}}, \bibinfo {author}
  {\bibfnamefont {J.}~\bibnamefont {Brucato}}, \bibinfo {author} {\bibfnamefont
  {H.}~\bibnamefont {Busemann}}, \bibinfo {author} {\bibfnamefont
  {L.}~\bibnamefont {Colangeli}}, \bibinfo {author} {\bibfnamefont
  {L.}~\bibnamefont {d'Hendecourt}}, \bibinfo {author} {\bibfnamefont
  {Z.}~\bibnamefont {Djouadi}}, \bibinfo {author} {\bibfnamefont
  {G.}~\bibnamefont {Ferrini}}, \bibinfo {author} {\bibfnamefont
  {I.}~\bibnamefont {Franchi}},  \emph {et~al.},\ }\bibfield  {title} {\enquote
  {\bibinfo {title} {Combined micro-raman, micro-infrared, and field emission
  scanning electron microscope analyses of comet 81p/wild 2 particles collected
  by stardust},}\ }\href@noop {} {\bibfield  {journal} {\bibinfo  {journal}
  {Meteoritics \& Planetary Science}\ }\textbf {\bibinfo {volume} {43}},\
  \bibinfo {pages} {367--397} (\bibinfo {year} {2008})}\BibitemShut {NoStop}%
\bibitem [{\citenamefont {Davidson}, \citenamefont {Busemann},\ and\
  \citenamefont {Franchi}(2012)}]{davidson2012nanosims}%
  \BibitemOpen
  \bibfield  {author} {\bibinfo {author} {\bibfnamefont {J.}~\bibnamefont
  {Davidson}}, \bibinfo {author} {\bibfnamefont {H.}~\bibnamefont {Busemann}},
  \ and\ \bibinfo {author} {\bibfnamefont {I.~A.}\ \bibnamefont {Franchi}},\
  }\bibfield  {title} {\enquote {\bibinfo {title} {A nanosims and raman
  spectroscopic comparison of interplanetary dust particles from comet
  grigg-skjellerup and non-grigg skjellerup collections},}\ }\href@noop {}
  {\bibfield  {journal} {\bibinfo  {journal} {Meteoritics \& Planetary
  Science}\ }\textbf {\bibinfo {volume} {47}},\ \bibinfo {pages} {1748--1771}
  (\bibinfo {year} {2012})}\BibitemShut {NoStop}%
\bibitem [{\citenamefont {Floss}\ \emph {et~al.}(2006)\citenamefont {Floss},
  \citenamefont {Stadermann}, \citenamefont {Bradley}, \citenamefont {Bajt},
  \citenamefont {Graham}, \citenamefont {Lea} \emph
  {et~al.}}]{floss2006identification}%
  \BibitemOpen
  \bibfield  {author} {\bibinfo {author} {\bibfnamefont {C.}~\bibnamefont
  {Floss}}, \bibinfo {author} {\bibfnamefont {F.~J.}\ \bibnamefont
  {Stadermann}}, \bibinfo {author} {\bibfnamefont {J.~P.}\ \bibnamefont
  {Bradley}}, \bibinfo {author} {\bibfnamefont {S.}~\bibnamefont {Bajt}},
  \bibinfo {author} {\bibfnamefont {G.}~\bibnamefont {Graham}}, \bibinfo
  {author} {\bibfnamefont {A.~S.}\ \bibnamefont {Lea}},  \emph {et~al.},\
  }\bibfield  {title} {\enquote {\bibinfo {title} {Identification of
  isotopically primitive interplanetary dust particles: A nanosims isotopic
  imaging study},}\ }\href@noop {} {\bibfield  {journal} {\bibinfo  {journal}
  {Geochimica et Cosmochimica Acta}\ }\textbf {\bibinfo {volume} {70}},\
  \bibinfo {pages} {2371--2399} (\bibinfo {year} {2006})}\BibitemShut {NoStop}%
\bibitem [{\citenamefont {Alali}\ \emph {et~al.}(2020)\citenamefont {Alali},
  \citenamefont {Gong}, \citenamefont {Videen}, \citenamefont {Pan},
  \citenamefont {Mu{\~n}oz},\ and\ \citenamefont {Wang}}]{alali2020laser}%
  \BibitemOpen
  \bibfield  {author} {\bibinfo {author} {\bibfnamefont {H.}~\bibnamefont
  {Alali}}, \bibinfo {author} {\bibfnamefont {Z.}~\bibnamefont {Gong}},
  \bibinfo {author} {\bibfnamefont {G.}~\bibnamefont {Videen}}, \bibinfo
  {author} {\bibfnamefont {Y.-L.}\ \bibnamefont {Pan}}, \bibinfo {author}
  {\bibfnamefont {O.}~\bibnamefont {Mu{\~n}oz}}, \ and\ \bibinfo {author}
  {\bibfnamefont {C.}~\bibnamefont {Wang}},\ }\bibfield  {title} {\enquote
  {\bibinfo {title} {Laser spectroscopic characterization of single
  extraterrestrial dust particles using optical trapping-cavity ringdown and
  raman spectroscopy},}\ }\href@noop {} {\bibfield  {journal} {\bibinfo
  {journal} {Journal of Quantitative Spectroscopy and Radiative Transfer}\
  }\textbf {\bibinfo {volume} {255}},\ \bibinfo {pages} {107249} (\bibinfo
  {year} {2020})}\BibitemShut {NoStop}%
\bibitem [{\citenamefont {Jones}, \citenamefont {Marag{\`o}},\ and\
  \citenamefont {Volpe}(2015)}]{jones2015optical}%
  \BibitemOpen
  \bibfield  {author} {\bibinfo {author} {\bibfnamefont {P.~H.}\ \bibnamefont
  {Jones}}, \bibinfo {author} {\bibfnamefont {O.~M.}\ \bibnamefont
  {Marag{\`o}}}, \ and\ \bibinfo {author} {\bibfnamefont {G.}~\bibnamefont
  {Volpe}},\ }\href@noop {} {\emph {\bibinfo {title} {Optical tweezers:
  Principles and applications}}}\ (\bibinfo  {publisher} {Cambridge University
  Press},\ \bibinfo {year} {2015})\BibitemShut {NoStop}%
\bibitem [{\citenamefont {Ashkin}\ and\ \citenamefont
  {Dziedzic}(1971)}]{ashkin1971optical}%
  \BibitemOpen
  \bibfield  {author} {\bibinfo {author} {\bibfnamefont {A.}~\bibnamefont
  {Ashkin}}\ and\ \bibinfo {author} {\bibfnamefont {J.~M.}\ \bibnamefont
  {Dziedzic}},\ }\bibfield  {title} {\enquote {\bibinfo {title} {Optical
  levitation by radiation pressure},}\ }\href {\doibase 10.1063/1.1653919}
  {\bibfield  {journal} {\bibinfo  {journal} {Applied Physics Letters}\
  }\textbf {\bibinfo {volume} {19}},\ \bibinfo {pages} {283--285} (\bibinfo
  {year} {1971})}\BibitemShut {NoStop}%
\bibitem [{\citenamefont {Ashkin}\ \emph {et~al.}(1986)\citenamefont {Ashkin},
  \citenamefont {Dziedzic}, \citenamefont {Bjorkholm},\ and\ \citenamefont
  {Chu}}]{ashkin1986observation}%
  \BibitemOpen
  \bibfield  {author} {\bibinfo {author} {\bibfnamefont {A.}~\bibnamefont
  {Ashkin}}, \bibinfo {author} {\bibfnamefont {J.~M.}\ \bibnamefont
  {Dziedzic}}, \bibinfo {author} {\bibfnamefont {J.~E.}\ \bibnamefont
  {Bjorkholm}}, \ and\ \bibinfo {author} {\bibfnamefont {S.}~\bibnamefont
  {Chu}},\ }\bibfield  {title} {\enquote {\bibinfo {title} {Observation of a
  single-beam gradient force optical trap for dielectric particles},}\
  }\href@noop {} {\bibfield  {journal} {\bibinfo  {journal} {Optics letters}\
  }\textbf {\bibinfo {volume} {11}},\ \bibinfo {pages} {288--290} (\bibinfo
  {year} {1986})}\BibitemShut {NoStop}%
\bibitem [{\citenamefont {Polimeno}\ \emph {et~al.}(2018)\citenamefont
  {Polimeno}, \citenamefont {Magazz{\`u}}, \citenamefont {Iati}, \citenamefont
  {Patti}, \citenamefont {Saija}, \citenamefont {Boschi}, \citenamefont
  {Donato}, \citenamefont {Gucciardi}, \citenamefont {Jones}, \citenamefont
  {Volpe} \emph {et~al.}}]{polimeno2018optical}%
  \BibitemOpen
  \bibfield  {author} {\bibinfo {author} {\bibfnamefont {P.}~\bibnamefont
  {Polimeno}}, \bibinfo {author} {\bibfnamefont {A.}~\bibnamefont
  {Magazz{\`u}}}, \bibinfo {author} {\bibfnamefont {M.~A.}\ \bibnamefont
  {Iati}}, \bibinfo {author} {\bibfnamefont {F.}~\bibnamefont {Patti}},
  \bibinfo {author} {\bibfnamefont {R.}~\bibnamefont {Saija}}, \bibinfo
  {author} {\bibfnamefont {C.~D.~E.}\ \bibnamefont {Boschi}}, \bibinfo {author}
  {\bibfnamefont {M.~G.}\ \bibnamefont {Donato}}, \bibinfo {author}
  {\bibfnamefont {P.~G.}\ \bibnamefont {Gucciardi}}, \bibinfo {author}
  {\bibfnamefont {P.~H.}\ \bibnamefont {Jones}}, \bibinfo {author}
  {\bibfnamefont {G.}~\bibnamefont {Volpe}},  \emph {et~al.},\ }\bibfield
  {title} {\enquote {\bibinfo {title} {Optical tweezers and their
  applications},}\ }\href@noop {} {\bibfield  {journal} {\bibinfo  {journal}
  {Journal of Quantitative Spectroscopy and Radiative Transfer}\ }\textbf
  {\bibinfo {volume} {218}},\ \bibinfo {pages} {131--150} (\bibinfo {year}
  {2018})}\BibitemShut {NoStop}%
\bibitem [{\citenamefont {Thurn}\ and\ \citenamefont
  {Kiefer}(1984)}]{thurn1984raman}%
  \BibitemOpen
  \bibfield  {author} {\bibinfo {author} {\bibfnamefont {R.}~\bibnamefont
  {Thurn}}\ and\ \bibinfo {author} {\bibfnamefont {W.}~\bibnamefont {Kiefer}},\
  }\bibfield  {title} {\enquote {\bibinfo {title} {Raman-microsampling
  technique applying optical levitation by radiation pressure},}\ }\href@noop
  {} {\bibfield  {journal} {\bibinfo  {journal} {Applied spectroscopy}\
  }\textbf {\bibinfo {volume} {38}},\ \bibinfo {pages} {78--83} (\bibinfo
  {year} {1984})}\BibitemShut {NoStop}%
\bibitem [{\citenamefont {Lankers}, \citenamefont {Popp},\ and\ \citenamefont
  {Kiefer}(1994)}]{lankers1994raman}%
  \BibitemOpen
  \bibfield  {author} {\bibinfo {author} {\bibfnamefont {M.}~\bibnamefont
  {Lankers}}, \bibinfo {author} {\bibfnamefont {J.}~\bibnamefont {Popp}}, \
  and\ \bibinfo {author} {\bibfnamefont {W.}~\bibnamefont {Kiefer}},\
  }\bibfield  {title} {\enquote {\bibinfo {title} {Raman and fluorescence
  spectra of single optically trapped microdroplets in emulsions},}\
  }\href@noop {} {\bibfield  {journal} {\bibinfo  {journal} {Applied
  spectroscopy}\ }\textbf {\bibinfo {volume} {48}},\ \bibinfo {pages}
  {1166--1168} (\bibinfo {year} {1994})}\BibitemShut {NoStop}%
\bibitem [{\citenamefont {Pan}, \citenamefont {Hill},\ and\ \citenamefont
  {Coleman}(2012)}]{pan2012photophoretic}%
  \BibitemOpen
  \bibfield  {author} {\bibinfo {author} {\bibfnamefont {Y.-L.}\ \bibnamefont
  {Pan}}, \bibinfo {author} {\bibfnamefont {S.~C.}\ \bibnamefont {Hill}}, \
  and\ \bibinfo {author} {\bibfnamefont {M.}~\bibnamefont {Coleman}},\
  }\bibfield  {title} {\enquote {\bibinfo {title} {Photophoretic trapping of
  absorbing particles in air and measurement of their single-particle raman
  spectra},}\ }\href@noop {} {\bibfield  {journal} {\bibinfo  {journal} {Optics
  express}\ }\textbf {\bibinfo {volume} {20}},\ \bibinfo {pages} {5325--5334}
  (\bibinfo {year} {2012})}\BibitemShut {NoStop}%
\bibitem [{\citenamefont {Gong}\ \emph {et~al.}(2018)\citenamefont {Gong},
  \citenamefont {Pan}, \citenamefont {Videen},\ and\ \citenamefont
  {Wang}}]{gong2018optical}%
  \BibitemOpen
  \bibfield  {author} {\bibinfo {author} {\bibfnamefont {Z.}~\bibnamefont
  {Gong}}, \bibinfo {author} {\bibfnamefont {Y.-L.}\ \bibnamefont {Pan}},
  \bibinfo {author} {\bibfnamefont {G.}~\bibnamefont {Videen}}, \ and\ \bibinfo
  {author} {\bibfnamefont {C.}~\bibnamefont {Wang}},\ }\bibfield  {title}
  {\enquote {\bibinfo {title} {Optical trapping and manipulation of single
  particles in air: Principles, technical details, and applications},}\
  }\href@noop {} {\bibfield  {journal} {\bibinfo  {journal} {Journal of
  Quantitative Spectroscopy and Radiative Transfer}\ }\textbf {\bibinfo
  {volume} {214}},\ \bibinfo {pages} {94--119} (\bibinfo {year}
  {2018})}\BibitemShut {NoStop}%
\bibitem [{\citenamefont {Gillibert}\ \emph {et~al.}(2019)\citenamefont
  {Gillibert}, \citenamefont {Balakrishnan}, \citenamefont {Deshoules},
  \citenamefont {Tardivel}, \citenamefont {Magazz{\`u}}, \citenamefont
  {Donato}, \citenamefont {Marag{\`o}}, \citenamefont {Lamy~de La~Chapelle},
  \citenamefont {Colas}, \citenamefont {Lagarde} \emph
  {et~al.}}]{gillibert2019raman}%
  \BibitemOpen
  \bibfield  {author} {\bibinfo {author} {\bibfnamefont {R.}~\bibnamefont
  {Gillibert}}, \bibinfo {author} {\bibfnamefont {G.}~\bibnamefont
  {Balakrishnan}}, \bibinfo {author} {\bibfnamefont {Q.}~\bibnamefont
  {Deshoules}}, \bibinfo {author} {\bibfnamefont {M.}~\bibnamefont {Tardivel}},
  \bibinfo {author} {\bibfnamefont {A.}~\bibnamefont {Magazz{\`u}}}, \bibinfo
  {author} {\bibfnamefont {M.~G.}\ \bibnamefont {Donato}}, \bibinfo {author}
  {\bibfnamefont {O.~M.}\ \bibnamefont {Marag{\`o}}}, \bibinfo {author}
  {\bibfnamefont {M.}~\bibnamefont {Lamy~de La~Chapelle}}, \bibinfo {author}
  {\bibfnamefont {F.}~\bibnamefont {Colas}}, \bibinfo {author} {\bibfnamefont
  {F.}~\bibnamefont {Lagarde}},  \emph {et~al.},\ }\bibfield  {title} {\enquote
  {\bibinfo {title} {Raman tweezers for small microplastics and nanoplastics
  identification in seawater},}\ }\href@noop {} {\bibfield  {journal} {\bibinfo
   {journal} {Environmental science \& technology}\ }\textbf {\bibinfo {volume}
  {53}},\ \bibinfo {pages} {9003--9013} (\bibinfo {year} {2019})}\BibitemShut
  {NoStop}%
\bibitem [{\citenamefont {Fazio}\ \emph {et~al.}(2014)\citenamefont {Fazio},
  \citenamefont {Folco}, \citenamefont {D'Orazio}, \citenamefont {Frezzotti},\
  and\ \citenamefont {Cordier}}]{fazio2014shock}%
  \BibitemOpen
  \bibfield  {author} {\bibinfo {author} {\bibfnamefont {A.}~\bibnamefont
  {Fazio}}, \bibinfo {author} {\bibfnamefont {L.}~\bibnamefont {Folco}},
  \bibinfo {author} {\bibfnamefont {M.}~\bibnamefont {D'Orazio}}, \bibinfo
  {author} {\bibfnamefont {M.~L.}\ \bibnamefont {Frezzotti}}, \ and\ \bibinfo
  {author} {\bibfnamefont {C.}~\bibnamefont {Cordier}},\ }\bibfield  {title}
  {\enquote {\bibinfo {title} {Shock metamorphism and impact melting in small
  impact craters on earth: Evidence from kamil crater, egypt},}\ }\href@noop {}
  {\bibfield  {journal} {\bibinfo  {journal} {Meteoritics \& Planetary
  Science}\ }\textbf {\bibinfo {volume} {49}},\ \bibinfo {pages} {2175--2200}
  (\bibinfo {year} {2014})}\BibitemShut {NoStop}%
\bibitem [{\citenamefont {Collareta}\ \emph {et~al.}(2016)\citenamefont
  {Collareta}, \citenamefont {D'Orazio}, \citenamefont {Gemelli}, \citenamefont
  {Pack},\ and\ \citenamefont {Folco}}]{collareta2016high}%
  \BibitemOpen
  \bibfield  {author} {\bibinfo {author} {\bibfnamefont {A.}~\bibnamefont
  {Collareta}}, \bibinfo {author} {\bibfnamefont {M.}~\bibnamefont {D'Orazio}},
  \bibinfo {author} {\bibfnamefont {M.}~\bibnamefont {Gemelli}}, \bibinfo
  {author} {\bibfnamefont {A.}~\bibnamefont {Pack}}, \ and\ \bibinfo {author}
  {\bibfnamefont {L.}~\bibnamefont {Folco}},\ }\bibfield  {title} {\enquote
  {\bibinfo {title} {High crustal diversity preserved in the lunar meteorite
  mount dewitt 12007 (victoria land, antarctica)},}\ }\href@noop {} {\bibfield
  {journal} {\bibinfo  {journal} {Meteoritics \& Planetary Science}\ }\textbf
  {\bibinfo {volume} {51}},\ \bibinfo {pages} {351--371} (\bibinfo {year}
  {2016})}\BibitemShut {NoStop}%
\bibitem [{\citenamefont {Colangeli}\ \emph {et~al.}(2007)\citenamefont
  {Colangeli}, \citenamefont {Lopez-Moreno}, \citenamefont {Palumbo},
  \citenamefont {Rodriguez}, \citenamefont {Cosi}, \citenamefont {Della~Corte},
  \citenamefont {Esposito}, \citenamefont {Fulle}, \citenamefont {Herranz},
  \citenamefont {Jeronimo} \emph {et~al.}}]{colangeli2007grain}%
  \BibitemOpen
  \bibfield  {author} {\bibinfo {author} {\bibfnamefont {L.}~\bibnamefont
  {Colangeli}}, \bibinfo {author} {\bibfnamefont {J.}~\bibnamefont
  {Lopez-Moreno}}, \bibinfo {author} {\bibfnamefont {P.}~\bibnamefont
  {Palumbo}}, \bibinfo {author} {\bibfnamefont {J.}~\bibnamefont {Rodriguez}},
  \bibinfo {author} {\bibfnamefont {M.}~\bibnamefont {Cosi}}, \bibinfo {author}
  {\bibfnamefont {V.}~\bibnamefont {Della~Corte}}, \bibinfo {author}
  {\bibfnamefont {F.}~\bibnamefont {Esposito}}, \bibinfo {author}
  {\bibfnamefont {M.}~\bibnamefont {Fulle}}, \bibinfo {author} {\bibfnamefont
  {M.}~\bibnamefont {Herranz}}, \bibinfo {author} {\bibfnamefont
  {J.}~\bibnamefont {Jeronimo}},  \emph {et~al.},\ }\bibfield  {title}
  {\enquote {\bibinfo {title} {The grain impact analyser and dust accumulator
  (giada) experiment for the rosetta mission: design, performances and first
  results},}\ }\href@noop {} {\bibfield  {journal} {\bibinfo  {journal} {Space
  Science Reviews}\ }\textbf {\bibinfo {volume} {128}},\ \bibinfo {pages}
  {803--821} (\bibinfo {year} {2007})}\BibitemShut {NoStop}%
\bibitem [{\citenamefont {Gieseler}\ \emph {et~al.}(2021)\citenamefont
  {Gieseler}, \citenamefont {Gomez-Solano}, \citenamefont {Magazz{\`u}},
  \citenamefont {Castillo}, \citenamefont {Garc{\'\i}a}, \citenamefont
  {Gironella-Torrent}, \citenamefont {Viader-Godoy}, \citenamefont {Ritort},
  \citenamefont {Pesce}, \citenamefont {Arzola} \emph
  {et~al.}}]{gieseler2021optical}%
  \BibitemOpen
  \bibfield  {author} {\bibinfo {author} {\bibfnamefont {J.}~\bibnamefont
  {Gieseler}}, \bibinfo {author} {\bibfnamefont {J.~R.}\ \bibnamefont
  {Gomez-Solano}}, \bibinfo {author} {\bibfnamefont {A.}~\bibnamefont
  {Magazz{\`u}}}, \bibinfo {author} {\bibfnamefont {I.~P.}\ \bibnamefont
  {Castillo}}, \bibinfo {author} {\bibfnamefont {L.~P.}\ \bibnamefont
  {Garc{\'\i}a}}, \bibinfo {author} {\bibfnamefont {M.}~\bibnamefont
  {Gironella-Torrent}}, \bibinfo {author} {\bibfnamefont {X.}~\bibnamefont
  {Viader-Godoy}}, \bibinfo {author} {\bibfnamefont {F.}~\bibnamefont
  {Ritort}}, \bibinfo {author} {\bibfnamefont {G.}~\bibnamefont {Pesce}},
  \bibinfo {author} {\bibfnamefont {A.~V.}\ \bibnamefont {Arzola}},  \emph
  {et~al.},\ }\bibfield  {title} {\enquote {\bibinfo {title} {Optical
  tweezers—from calibration to applications: a tutorial},}\ }\href@noop {}
  {\bibfield  {journal} {\bibinfo  {journal} {Advances in Optics and
  Photonics}\ }\textbf {\bibinfo {volume} {13}},\ \bibinfo {pages} {74--241}
  (\bibinfo {year} {2021})}\BibitemShut {NoStop}%
\bibitem [{\citenamefont {Borghese}, \citenamefont {Denti},\ and\ \citenamefont
  {Saija}(2007{\natexlab{a}})}]{borghese2007scattering}%
  \BibitemOpen
  \bibfield  {author} {\bibinfo {author} {\bibfnamefont {F.}~\bibnamefont
  {Borghese}}, \bibinfo {author} {\bibfnamefont {P.}~\bibnamefont {Denti}}, \
  and\ \bibinfo {author} {\bibfnamefont {R.}~\bibnamefont {Saija}},\
  }\href@noop {} {\emph {\bibinfo {title} {Scattering from model nonspherical
  particles: theory and applications to environmental physics}}}\ (\bibinfo
  {publisher} {Springer Science \& Business Media},\ \bibinfo {year}
  {2007})\BibitemShut {NoStop}%
\bibitem [{\citenamefont {Polimeno}\ \emph {et~al.}(2021)\citenamefont
  {Polimeno}, \citenamefont {Magazz{\`u}}, \citenamefont {Iat{\`\i}},
  \citenamefont {Saija}, \citenamefont {Folco}, \citenamefont {Ciriza},
  \citenamefont {Donato}, \citenamefont {Foti}, \citenamefont {Gucciardi},
  \citenamefont {Saidi} \emph {et~al.}}]{polimeno2021optical}%
  \BibitemOpen
  \bibfield  {author} {\bibinfo {author} {\bibfnamefont {P.}~\bibnamefont
  {Polimeno}}, \bibinfo {author} {\bibfnamefont {A.}~\bibnamefont
  {Magazz{\`u}}}, \bibinfo {author} {\bibfnamefont {M.}~\bibnamefont
  {Iat{\`\i}}}, \bibinfo {author} {\bibfnamefont {R.}~\bibnamefont {Saija}},
  \bibinfo {author} {\bibfnamefont {L.}~\bibnamefont {Folco}}, \bibinfo
  {author} {\bibfnamefont {D.~B.}\ \bibnamefont {Ciriza}}, \bibinfo {author}
  {\bibfnamefont {M.}~\bibnamefont {Donato}}, \bibinfo {author} {\bibfnamefont
  {A.}~\bibnamefont {Foti}}, \bibinfo {author} {\bibfnamefont {P.}~\bibnamefont
  {Gucciardi}}, \bibinfo {author} {\bibfnamefont {A.}~\bibnamefont {Saidi}},
  \emph {et~al.},\ }\bibfield  {title} {\enquote {\bibinfo {title} {Optical
  tweezers in a dusty universe},}\ }\href@noop {} {\bibfield  {journal}
  {\bibinfo  {journal} {The European Physical Journal Plus}\ }\textbf {\bibinfo
  {volume} {136}},\ \bibinfo {pages} {1--23} (\bibinfo {year}
  {2021})}\BibitemShut {NoStop}%
\bibitem [{\citenamefont {Bohren}\ and\ \citenamefont
  {Huffman}(2008)}]{bohren2008absorption}%
  \BibitemOpen
  \bibfield  {author} {\bibinfo {author} {\bibfnamefont {C.~F.}\ \bibnamefont
  {Bohren}}\ and\ \bibinfo {author} {\bibfnamefont {D.~R.}\ \bibnamefont
  {Huffman}},\ }\href@noop {} {\emph {\bibinfo {title} {Absorption and
  scattering of light by small particles}}}\ (\bibinfo  {publisher} {John Wiley
  \& Sons},\ \bibinfo {year} {2008})\BibitemShut {NoStop}%
\bibitem [{\citenamefont {Marston}\ and\ \citenamefont
  {Crichton}(1984)}]{marston1984radiation}%
  \BibitemOpen
  \bibfield  {author} {\bibinfo {author} {\bibfnamefont {P.~L.}\ \bibnamefont
  {Marston}}\ and\ \bibinfo {author} {\bibfnamefont {J.~H.}\ \bibnamefont
  {Crichton}},\ }\bibfield  {title} {\enquote {\bibinfo {title} {Radiation
  torque on a sphere caused by a circularly-polarized electromagnetic wave},}\
  }\href@noop {} {\bibfield  {journal} {\bibinfo  {journal} {Physical Review
  A}\ }\textbf {\bibinfo {volume} {30}},\ \bibinfo {pages} {2508} (\bibinfo
  {year} {1984})}\BibitemShut {NoStop}%
\bibitem [{\citenamefont {Swartzlander}\ \emph {et~al.}(2011)\citenamefont
  {Swartzlander}, \citenamefont {Peterson}, \citenamefont {Artusio-Glimpse},\
  and\ \citenamefont {Raisanen}}]{swartzlander2011stable}%
  \BibitemOpen
  \bibfield  {author} {\bibinfo {author} {\bibfnamefont {G.~A.}\ \bibnamefont
  {Swartzlander}}, \bibinfo {author} {\bibfnamefont {T.~J.}\ \bibnamefont
  {Peterson}}, \bibinfo {author} {\bibfnamefont {A.~B.}\ \bibnamefont
  {Artusio-Glimpse}}, \ and\ \bibinfo {author} {\bibfnamefont {A.~D.}\
  \bibnamefont {Raisanen}},\ }\bibfield  {title} {\enquote {\bibinfo {title}
  {Stable optical lift},}\ }\href@noop {} {\bibfield  {journal} {\bibinfo
  {journal} {Nature Photonics}\ }\textbf {\bibinfo {volume} {5}},\ \bibinfo
  {pages} {48--51} (\bibinfo {year} {2011})}\BibitemShut {NoStop}%
\bibitem [{\citenamefont {Irrera}\ \emph {et~al.}(2016)\citenamefont {Irrera},
  \citenamefont {Magazz{\`u}}, \citenamefont {Artoni}, \citenamefont {Simpson},
  \citenamefont {Hanna}, \citenamefont {Jones}, \citenamefont {Priolo},
  \citenamefont {Gucciardi},\ and\ \citenamefont
  {Marag{\`o}}}]{irrera2016photonic}%
  \BibitemOpen
  \bibfield  {author} {\bibinfo {author} {\bibfnamefont {A.}~\bibnamefont
  {Irrera}}, \bibinfo {author} {\bibfnamefont {A.}~\bibnamefont {Magazz{\`u}}},
  \bibinfo {author} {\bibfnamefont {P.}~\bibnamefont {Artoni}}, \bibinfo
  {author} {\bibfnamefont {S.~H.}\ \bibnamefont {Simpson}}, \bibinfo {author}
  {\bibfnamefont {S.}~\bibnamefont {Hanna}}, \bibinfo {author} {\bibfnamefont
  {P.~H.}\ \bibnamefont {Jones}}, \bibinfo {author} {\bibfnamefont
  {F.}~\bibnamefont {Priolo}}, \bibinfo {author} {\bibfnamefont {P.~G.}\
  \bibnamefont {Gucciardi}}, \ and\ \bibinfo {author} {\bibfnamefont {O.~M.}\
  \bibnamefont {Marag{\`o}}},\ }\bibfield  {title} {\enquote {\bibinfo {title}
  {Photonic torque microscopy of the nonconservative force field for optically
  trapped silicon nanowires},}\ }\href@noop {} {\bibfield  {journal} {\bibinfo
  {journal} {Nano letters}\ }\textbf {\bibinfo {volume} {16}},\ \bibinfo
  {pages} {4181--4188} (\bibinfo {year} {2016})}\BibitemShut {NoStop}%
\bibitem [{\citenamefont {Schmidt}\ \emph {et~al.}(2018)\citenamefont
  {Schmidt}, \citenamefont {Magazz{\`u}}, \citenamefont {Callegari},
  \citenamefont {Biancofiore}, \citenamefont {Cichos},\ and\ \citenamefont
  {Volpe}}]{schmidt2018microscopic}%
  \BibitemOpen
  \bibfield  {author} {\bibinfo {author} {\bibfnamefont {F.}~\bibnamefont
  {Schmidt}}, \bibinfo {author} {\bibfnamefont {A.}~\bibnamefont
  {Magazz{\`u}}}, \bibinfo {author} {\bibfnamefont {A.}~\bibnamefont
  {Callegari}}, \bibinfo {author} {\bibfnamefont {L.}~\bibnamefont
  {Biancofiore}}, \bibinfo {author} {\bibfnamefont {F.}~\bibnamefont {Cichos}},
  \ and\ \bibinfo {author} {\bibfnamefont {G.}~\bibnamefont {Volpe}},\
  }\bibfield  {title} {\enquote {\bibinfo {title} {Microscopic engine powered
  by critical demixing},}\ }\href@noop {} {\bibfield  {journal} {\bibinfo
  {journal} {Physical Review Letters}\ }\textbf {\bibinfo {volume} {120}},\
  \bibinfo {pages} {068004} (\bibinfo {year} {2018})}\BibitemShut {NoStop}%
\bibitem [{\citenamefont {Pesce}\ \emph {et~al.}(2009)\citenamefont {Pesce},
  \citenamefont {Volpe}, \citenamefont {De~Luca}, \citenamefont {Rusciano},\
  and\ \citenamefont {Volpe}}]{pesce2009quantitative}%
  \BibitemOpen
  \bibfield  {author} {\bibinfo {author} {\bibfnamefont {G.}~\bibnamefont
  {Pesce}}, \bibinfo {author} {\bibfnamefont {G.}~\bibnamefont {Volpe}},
  \bibinfo {author} {\bibfnamefont {A.~C.}\ \bibnamefont {De~Luca}}, \bibinfo
  {author} {\bibfnamefont {G.}~\bibnamefont {Rusciano}}, \ and\ \bibinfo
  {author} {\bibfnamefont {G.}~\bibnamefont {Volpe}},\ }\bibfield  {title}
  {\enquote {\bibinfo {title} {Quantitative assessment of non-conservative
  radiation forces in an optical trap},}\ }\href@noop {} {\bibfield  {journal}
  {\bibinfo  {journal} {EPL (Europhysics Letters)}\ }\textbf {\bibinfo {volume}
  {86}},\ \bibinfo {pages} {38002} (\bibinfo {year} {2009})}\BibitemShut
  {NoStop}%
\bibitem [{\citenamefont {Jones}\ \emph {et~al.}(2009)\citenamefont {Jones},
  \citenamefont {Palmisano}, \citenamefont {Bonaccorso}, \citenamefont
  {Gucciardi}, \citenamefont {Calogero}, \citenamefont {Ferrari},\ and\
  \citenamefont {Marag{\`o}}}]{jones2009rotation}%
  \BibitemOpen
  \bibfield  {author} {\bibinfo {author} {\bibfnamefont {P.}~\bibnamefont
  {Jones}}, \bibinfo {author} {\bibfnamefont {F.}~\bibnamefont {Palmisano}},
  \bibinfo {author} {\bibfnamefont {F.}~\bibnamefont {Bonaccorso}}, \bibinfo
  {author} {\bibfnamefont {P.}~\bibnamefont {Gucciardi}}, \bibinfo {author}
  {\bibfnamefont {G.}~\bibnamefont {Calogero}}, \bibinfo {author}
  {\bibfnamefont {A.}~\bibnamefont {Ferrari}}, \ and\ \bibinfo {author}
  {\bibfnamefont {O.~M.}\ \bibnamefont {Marag{\`o}}},\ }\bibfield  {title}
  {\enquote {\bibinfo {title} {Rotation detection in light-driven
  nanorotors},}\ }\href@noop {} {\bibfield  {journal} {\bibinfo  {journal} {ACS
  nano}\ }\textbf {\bibinfo {volume} {3}},\ \bibinfo {pages} {3077--3084}
  (\bibinfo {year} {2009})}\BibitemShut {NoStop}%
\bibitem [{\citenamefont {Orlando}\ \emph {et~al.}(2008)\citenamefont
  {Orlando}, \citenamefont {D’Orazio}, \citenamefont {Armienti},\ and\
  \citenamefont {Borrini}}]{orlando2008experimental}%
  \BibitemOpen
  \bibfield  {author} {\bibinfo {author} {\bibfnamefont {A.}~\bibnamefont
  {Orlando}}, \bibinfo {author} {\bibfnamefont {M.}~\bibnamefont {D’Orazio}},
  \bibinfo {author} {\bibfnamefont {P.}~\bibnamefont {Armienti}}, \ and\
  \bibinfo {author} {\bibfnamefont {D.}~\bibnamefont {Borrini}},\ }\bibfield
  {title} {\enquote {\bibinfo {title} {Experimental determination of
  plagioclase and clinopyroxene crystal growth rates in an anhydrous
  trachybasalt from mt etna (italy)},}\ }\href@noop {} {\bibfield  {journal}
  {\bibinfo  {journal} {European Journal of Mineralogy}\ }\textbf {\bibinfo
  {volume} {20}},\ \bibinfo {pages} {653--664} (\bibinfo {year}
  {2008})}\BibitemShut {NoStop}%
\bibitem [{\citenamefont {Brunetto}\ \emph {et~al.}(2014)\citenamefont
  {Brunetto}, \citenamefont {Lantz}, \citenamefont {Ledu}, \citenamefont
  {Baklouti}, \citenamefont {Barucci}, \citenamefont {Beck}, \citenamefont
  {Delauche}, \citenamefont {Dionnet}, \citenamefont {Dumas}, \citenamefont
  {Duprat} \emph {et~al.}}]{brunetto2014ion}%
  \BibitemOpen
  \bibfield  {author} {\bibinfo {author} {\bibfnamefont {R.}~\bibnamefont
  {Brunetto}}, \bibinfo {author} {\bibfnamefont {C.}~\bibnamefont {Lantz}},
  \bibinfo {author} {\bibfnamefont {D.}~\bibnamefont {Ledu}}, \bibinfo {author}
  {\bibfnamefont {D.}~\bibnamefont {Baklouti}}, \bibinfo {author}
  {\bibfnamefont {M.}~\bibnamefont {Barucci}}, \bibinfo {author} {\bibfnamefont
  {P.}~\bibnamefont {Beck}}, \bibinfo {author} {\bibfnamefont {L.}~\bibnamefont
  {Delauche}}, \bibinfo {author} {\bibfnamefont {Z.}~\bibnamefont {Dionnet}},
  \bibinfo {author} {\bibfnamefont {P.}~\bibnamefont {Dumas}}, \bibinfo
  {author} {\bibfnamefont {J.}~\bibnamefont {Duprat}},  \emph {et~al.},\
  }\bibfield  {title} {\enquote {\bibinfo {title} {Ion irradiation of allende
  meteorite probed by visible, ir, and raman spectroscopies},}\ }\href@noop {}
  {\bibfield  {journal} {\bibinfo  {journal} {Icarus}\ }\textbf {\bibinfo
  {volume} {237}},\ \bibinfo {pages} {278--292} (\bibinfo {year}
  {2014})}\BibitemShut {NoStop}%
\bibitem [{\citenamefont {Cristofolini}\ \emph {et~al.}(1987)\citenamefont
  {Cristofolini}, \citenamefont {Menzies}, \citenamefont {Beccaluva},\ and\
  \citenamefont {Tindle}}]{cristofolini1987petrological}%
  \BibitemOpen
  \bibfield  {author} {\bibinfo {author} {\bibfnamefont {R.}~\bibnamefont
  {Cristofolini}}, \bibinfo {author} {\bibfnamefont {M.}~\bibnamefont
  {Menzies}}, \bibinfo {author} {\bibfnamefont {L.}~\bibnamefont {Beccaluva}},
  \ and\ \bibinfo {author} {\bibfnamefont {A.}~\bibnamefont {Tindle}},\
  }\bibfield  {title} {\enquote {\bibinfo {title} {Petrological notes on the
  1983 lavas at mount etna, sicily, with reference to their ree and sr—nd
  isotope composition},}\ }\href@noop {} {\bibfield  {journal} {\bibinfo
  {journal} {Bulletin of volcanology}\ }\textbf {\bibinfo {volume} {49}},\
  \bibinfo {pages} {599--607} (\bibinfo {year} {1987})}\BibitemShut {NoStop}%
\bibitem [{\citenamefont {Smith}\ \emph {et~al.}(2021)\citenamefont {Smith},
  \citenamefont {Russell}, \citenamefont {Hutzler}, \citenamefont {Meneghin},
  \citenamefont {Brucato}, \citenamefont {Rettberg}, \citenamefont {Leuko},
  \citenamefont {Longobardo}, \citenamefont {Dirri}, \citenamefont {Palomba}
  \emph {et~al.}}]{smith2021roadmap}%
  \BibitemOpen
  \bibfield  {author} {\bibinfo {author} {\bibfnamefont {C.~L.}\ \bibnamefont
  {Smith}}, \bibinfo {author} {\bibfnamefont {S.~S.}\ \bibnamefont {Russell}},
  \bibinfo {author} {\bibfnamefont {A.}~\bibnamefont {Hutzler}}, \bibinfo
  {author} {\bibfnamefont {A.}~\bibnamefont {Meneghin}}, \bibinfo {author}
  {\bibfnamefont {J.~R.}\ \bibnamefont {Brucato}}, \bibinfo {author}
  {\bibfnamefont {P.}~\bibnamefont {Rettberg}}, \bibinfo {author}
  {\bibfnamefont {S.}~\bibnamefont {Leuko}}, \bibinfo {author} {\bibfnamefont
  {A.}~\bibnamefont {Longobardo}}, \bibinfo {author} {\bibfnamefont
  {F.}~\bibnamefont {Dirri}}, \bibinfo {author} {\bibfnamefont
  {E.}~\bibnamefont {Palomba}},  \emph {et~al.},\ }\bibfield  {title} {\enquote
  {\bibinfo {title} {A roadmap for a european extraterrestrial sample curation
  facility--the eurocares project},}\ }in\ \href@noop {} {\emph {\bibinfo
  {booktitle} {Sample Return Missions}}}\ (\bibinfo  {publisher} {Elsevier},\
  \bibinfo {year} {2021})\ pp.\ \bibinfo {pages} {249--268}\BibitemShut
  {NoStop}%
\bibitem [{\citenamefont {Magazz{\'u}}\ \emph {et~al.}(2015)\citenamefont
  {Magazz{\'u}}, \citenamefont {Spadaro}, \citenamefont {Donato}, \citenamefont
  {Sayed}, \citenamefont {Messina}, \citenamefont {D’Andrea}, \citenamefont
  {Foti}, \citenamefont {Fazio}, \citenamefont {Iat{\'\i}}, \citenamefont
  {Irrera} \emph {et~al.}}]{magazzu2015optical}%
  \BibitemOpen
  \bibfield  {author} {\bibinfo {author} {\bibfnamefont {A.}~\bibnamefont
  {Magazz{\'u}}}, \bibinfo {author} {\bibfnamefont {D.}~\bibnamefont
  {Spadaro}}, \bibinfo {author} {\bibfnamefont {M.~G.}\ \bibnamefont {Donato}},
  \bibinfo {author} {\bibfnamefont {R.}~\bibnamefont {Sayed}}, \bibinfo
  {author} {\bibfnamefont {E.}~\bibnamefont {Messina}}, \bibinfo {author}
  {\bibfnamefont {C.}~\bibnamefont {D’Andrea}}, \bibinfo {author}
  {\bibfnamefont {A.}~\bibnamefont {Foti}}, \bibinfo {author} {\bibfnamefont
  {B.}~\bibnamefont {Fazio}}, \bibinfo {author} {\bibfnamefont
  {M.}~\bibnamefont {Iat{\'\i}}}, \bibinfo {author} {\bibfnamefont
  {A.}~\bibnamefont {Irrera}},  \emph {et~al.},\ }\bibfield  {title} {\enquote
  {\bibinfo {title} {Optical tweezers: a non-destructive tool for soft and
  biomaterial investigations},}\ }\href@noop {} {\bibfield  {journal} {\bibinfo
   {journal} {Rendiconti Lincei}\ }\textbf {\bibinfo {volume} {26}},\ \bibinfo
  {pages} {203--218} (\bibinfo {year} {2015})}\BibitemShut {NoStop}%
\bibitem [{\citenamefont {Donato}\ \emph {et~al.}(2014)\citenamefont {Donato},
  \citenamefont {Hernandez}, \citenamefont {Mazzulla}, \citenamefont
  {Provenzano}, \citenamefont {Saija}, \citenamefont {Sayed}, \citenamefont
  {Vasi}, \citenamefont {Magazz{\`u}}, \citenamefont {Pagliusi}, \citenamefont
  {Bartolino} \emph {et~al.}}]{donato2014polarization}%
  \BibitemOpen
  \bibfield  {author} {\bibinfo {author} {\bibfnamefont {M.~G.}\ \bibnamefont
  {Donato}}, \bibinfo {author} {\bibfnamefont {J.}~\bibnamefont {Hernandez}},
  \bibinfo {author} {\bibfnamefont {A.}~\bibnamefont {Mazzulla}}, \bibinfo
  {author} {\bibfnamefont {C.}~\bibnamefont {Provenzano}}, \bibinfo {author}
  {\bibfnamefont {R.}~\bibnamefont {Saija}}, \bibinfo {author} {\bibfnamefont
  {R.}~\bibnamefont {Sayed}}, \bibinfo {author} {\bibfnamefont
  {S.}~\bibnamefont {Vasi}}, \bibinfo {author} {\bibfnamefont {A.}~\bibnamefont
  {Magazz{\`u}}}, \bibinfo {author} {\bibfnamefont {P.}~\bibnamefont
  {Pagliusi}}, \bibinfo {author} {\bibfnamefont {R.}~\bibnamefont {Bartolino}},
   \emph {et~al.},\ }\bibfield  {title} {\enquote {\bibinfo {title}
  {Polarization-dependent optomechanics mediated by chiral microresonators},}\
  }\href@noop {} {\bibfield  {journal} {\bibinfo  {journal} {Nature
  communications}\ }\textbf {\bibinfo {volume} {5}},\ \bibinfo {pages} {1--7}
  (\bibinfo {year} {2014})}\BibitemShut {NoStop}%
\bibitem [{\citenamefont {Gittes}\ and\ \citenamefont
  {Schmidt}(1998)}]{gittes1998interference}%
  \BibitemOpen
  \bibfield  {author} {\bibinfo {author} {\bibfnamefont {F.}~\bibnamefont
  {Gittes}}\ and\ \bibinfo {author} {\bibfnamefont {C.~F.}\ \bibnamefont
  {Schmidt}},\ }\bibfield  {title} {\enquote {\bibinfo {title} {Interference
  model for back-focal-plane displacement detection in optical tweezers},}\
  }\href@noop {} {\bibfield  {journal} {\bibinfo  {journal} {Optics letters}\
  }\textbf {\bibinfo {volume} {23}},\ \bibinfo {pages} {7--9} (\bibinfo {year}
  {1998})}\BibitemShut {NoStop}%
\bibitem [{\citenamefont {Pfeifer}\ \emph {et~al.}(2007)\citenamefont
  {Pfeifer}, \citenamefont {Nieminen}, \citenamefont {Heckenberg},\ and\
  \citenamefont {Rubinsztein-Dunlop}}]{pfeifer2007}%
  \BibitemOpen
  \bibfield  {author} {\bibinfo {author} {\bibfnamefont {R.~N.~C.}\
  \bibnamefont {Pfeifer}}, \bibinfo {author} {\bibfnamefont {T.~A.}\
  \bibnamefont {Nieminen}}, \bibinfo {author} {\bibfnamefont {N.~R.}\
  \bibnamefont {Heckenberg}}, \ and\ \bibinfo {author} {\bibfnamefont
  {H.}~\bibnamefont {Rubinsztein-Dunlop}},\ }\bibfield  {title} {\enquote
  {\bibinfo {title} {Colloquium: Momentum of an electromagnetic wave in
  dielectric media},}\ }\href@noop {} {\bibfield  {journal} {\bibinfo
  {journal} {Reviews of Modern Physics}\ }\textbf {\bibinfo {volume} {79}},\
  \bibinfo {pages} {1197} (\bibinfo {year} {2007})}\BibitemShut {NoStop}%
\bibitem [{\citenamefont {Saija}(2005)}]{saija2005transverse}%
  \BibitemOpen
  \bibfield  {author} {\bibinfo {author} {\bibfnamefont {R.~e.~a.}\
  \bibnamefont {Saija}},\ }\bibfield  {title} {\enquote {\bibinfo {title}
  {Transverse components of the radiation force on nonspherical particles in
  the t-matrix formalism},}\ }\href {\doibase 10.1016/j.jqsrt.2004.09.006}
  {\bibfield  {journal} {\bibinfo  {journal} {Journal of Quantitative
  Spectroscopy \& Radiative Transfer}\ }\textbf {\bibinfo {volume} {94}},\
  \bibinfo {pages} {163--179} (\bibinfo {year} {2005})}\BibitemShut {NoStop}%
\bibitem [{\citenamefont {Borghese}, \citenamefont {Denti},\ and\ \citenamefont
  {Saija}(2007{\natexlab{b}})}]{borghese2007optical}%
  \BibitemOpen
  \bibfield  {author} {\bibinfo {author} {\bibfnamefont {F.}~\bibnamefont
  {Borghese}}, \bibinfo {author} {\bibfnamefont {P.}~\bibnamefont {Denti}}, \
  and\ \bibinfo {author} {\bibfnamefont {R.}~\bibnamefont {Saija}},\ }\bibfield
   {title} {\enquote {\bibinfo {title} {Optical trapping of nonspherical
  particles in the t-matrix formalism},}\ }\href@noop {} {\bibfield  {journal}
  {\bibinfo  {journal} {Optics Express}\ }\textbf {\bibinfo {volume} {15}},\
  \bibinfo {pages} {11984--11998} (\bibinfo {year}
  {2007}{\natexlab{b}})}\BibitemShut {NoStop}%
\bibitem [{\citenamefont {Richards}(1959)}]{richard1959}%
  \BibitemOpen
  \bibfield  {author} {\bibinfo {author} {\bibfnamefont {E.}~\bibnamefont
  {Richards}, \bibfnamefont {B.~\&~Wolf}},\ }\bibfield  {title} {\enquote
  {\bibinfo {title} {Electromagnetic diffraction in optical systems. ii.
  structure of the image field in an aplanatic system.}}\ }\href@noop {}
  {\bibfield  {journal} {\bibinfo  {journal} {Proc. R. Soc. A: Math. Phys. Eng.
  Sci}\ }\textbf {\bibinfo {volume} {253}},\ \bibinfo {pages} {358} (\bibinfo
  {year} {1959})}\BibitemShut {NoStop}%
\bibitem [{\citenamefont {Garnett}(1904)}]{garnett1904}%
  \BibitemOpen
  \bibfield  {author} {\bibinfo {author} {\bibfnamefont {J.~C.~M.}\
  \bibnamefont {Garnett}},\ }\bibfield  {title} {\enquote {\bibinfo {title}
  {Colours in metal glasses and in metallic films i},}\ }\href@noop {}
  {\bibfield  {journal} {\bibinfo  {journal} {Philos. Trans. R. Soc. London}\
  }\textbf {\bibinfo {volume} {A 203}},\ \bibinfo {pages} {385--420} (\bibinfo
  {year} {1904})}\BibitemShut {NoStop}%
\bibitem [{\citenamefont {Garnett}(1906)}]{garnett1906}%
  \BibitemOpen
  \bibfield  {author} {\bibinfo {author} {\bibfnamefont {J.~C.~M.}\
  \bibnamefont {Garnett}},\ }\bibfield  {title} {\enquote {\bibinfo {title}
  {Colours in metal glasses, in metallic films, and in metallic solutions
  ii},}\ }\href@noop {} {\bibfield  {journal} {\bibinfo  {journal} {Philos.
  Trans. R. Soc. London}\ }\textbf {\bibinfo {volume} {A 205}},\ \bibinfo
  {pages} {237–288} (\bibinfo {year} {1906})}\BibitemShut {NoStop}%
\bibitem [{\citenamefont {Garnett}(1935)}]{bruggeman1935}%
  \BibitemOpen
  \bibfield  {author} {\bibinfo {author} {\bibfnamefont {J.~C.~M.}\
  \bibnamefont {Garnett}},\ }\bibfield  {title} {\enquote {\bibinfo {title}
  {Berechnung verschiedener physikalischer konstanten von heterogenen
  substanzen. i. dielektrizitätskonstanten und leitfähigkeiten der
  mischkörper aus isotropen substanzen},}\ }\href@noop {} {\bibfield
  {journal} {\bibinfo  {journal} {Ann. Phys.}\ }\textbf {\bibinfo {volume}
  {416}},\ \bibinfo {pages} {665–679} (\bibinfo {year} {1935})}\BibitemShut
  {NoStop}%
\bibitem [{\citenamefont {Garnett}(1936)}]{bruggeman1936}%
  \BibitemOpen
  \bibfield  {author} {\bibinfo {author} {\bibfnamefont {J.~C.~M.}\
  \bibnamefont {Garnett}},\ }\bibfield  {title} {\enquote {\bibinfo {title}
  {Berechnung verschiedener physikalischer konstanten von heterogenen
  substanzen. ii. dielektrizitätskonstanten und leitfähigkeiten von
  vielrkistallen der nichtregularen systeme},}\ }\href@noop {} {\bibfield
  {journal} {\bibinfo  {journal} {Ann. Phys.}\ }\textbf {\bibinfo {volume}
  {417}},\ \bibinfo {pages} {645–672} (\bibinfo {year} {1936})}\BibitemShut
  {NoStop}%
\bibitem [{\citenamefont {Borghese}, \citenamefont {Denti},\ and\ \citenamefont
  {Saija}(1994)}]{borghese1994eccentric}%
  \BibitemOpen
  \bibfield  {author} {\bibinfo {author} {\bibfnamefont {F.}~\bibnamefont
  {Borghese}}, \bibinfo {author} {\bibfnamefont {P.}~\bibnamefont {Denti}}, \
  and\ \bibinfo {author} {\bibfnamefont {R.}~\bibnamefont {Saija}},\ }\bibfield
   {title} {\enquote {\bibinfo {title} {Optical-properties of spheres
  containing several spherical inclusions},}\ }\href@noop {} {\bibfield
  {journal} {\bibinfo  {journal} {APPLIED OPTICS}\ }\textbf {\bibinfo {volume}
  {33}},\ \bibinfo {pages} {484--493} (\bibinfo {year} {1994})}\BibitemShut
  {NoStop}%
\bibitem [{\citenamefont {Borghese}, \citenamefont {Denti},\ and\ \citenamefont
  {Saija}(1998)}]{borghese1998eccentric}%
  \BibitemOpen
  \bibfield  {author} {\bibinfo {author} {\bibfnamefont {F.}~\bibnamefont
  {Borghese}}, \bibinfo {author} {\bibfnamefont {P.}~\bibnamefont {Denti}}, \
  and\ \bibinfo {author} {\bibfnamefont {R.}~\bibnamefont {Saija}},\ }\bibfield
   {title} {\enquote {\bibinfo {title} {Optical resonances of spheres
  containing an eccentric spherical inclusion},}\ }\href@noop {} {\bibfield
  {journal} {\bibinfo  {journal} {Journal of Optics}\ }\textbf {\bibinfo
  {volume} {29}},\ \bibinfo {pages} {28--34} (\bibinfo {year}
  {1998})}\BibitemShut {NoStop}%
\bibitem [{\citenamefont {Bish}\ and\ \citenamefont
  {Post}(1993)}]{bish1993quantitative}%
  \BibitemOpen
  \bibfield  {author} {\bibinfo {author} {\bibfnamefont {D.~L.}\ \bibnamefont
  {Bish}}\ and\ \bibinfo {author} {\bibfnamefont {J.~E.}\ \bibnamefont
  {Post}},\ }\bibfield  {title} {\enquote {\bibinfo {title} {Quantitative
  mineralogical analysis using the rietveld full-pattern fitting method},}\
  }\href@noop {} {\bibfield  {journal} {\bibinfo  {journal} {American
  Mineralogist}\ }\textbf {\bibinfo {volume} {78}},\ \bibinfo {pages}
  {932--940} (\bibinfo {year} {1993})}\BibitemShut {NoStop}%
\bibitem [{\citenamefont {Gualtieri}(2000)}]{gualtieri2000accuracy}%
  \BibitemOpen
  \bibfield  {author} {\bibinfo {author} {\bibfnamefont {A.~F.}\ \bibnamefont
  {Gualtieri}},\ }\bibfield  {title} {\enquote {\bibinfo {title} {Accuracy of
  xrpd qpa using the combined rietveld{--}rir method},}\ }\href@noop {}
  {\bibfield  {journal} {\bibinfo  {journal} {Journal of Applied
  Crystallography}\ }\textbf {\bibinfo {volume} {33}},\ \bibinfo {pages}
  {267--278} (\bibinfo {year} {2000})}\BibitemShut {NoStop}%
\end{thebibliography}%
%\bibliographystyle{aasjournal}



\end{document}